\documentclass[fleqn,12pt]{article}
\usepackage{amsfonts,epsfig,latexsym}
\usepackage{amssymb}
\usepackage{graphicx}
\usepackage{amsmath}
\usepackage{amssymb}
\usepackage{amsfonts}
\usepackage{graphicx}
\usepackage{mathrsfs}
\usepackage{xcolor}

\newcommand{\nnb}{\nonumber}
\newcommand{\pd}[2]{\frac{\partial #1}{\partial #2}}
\newcommand{\ep}{\varepsilon}

\newcommand{\om}{\omega}
\newcommand{\Om}{\Omega}
\newcommand{\re}{\mathrm{Re}}
\newcommand{\im}{\mathrm{Im}}
\newcommand{\cl}[1]{\mathcal{#1}}
\newcommand{\bb}[1]{\mathbb{#1}}
\newcommand{\ul}{\underline}
\newcommand{\ol}{\overline}

\newcommand{\ft}[2]{{\textstyle {\frac{#1}{#2}} }}

       \let\C=\Chi    

\newcommand{\C}{{\mathbb C}}

\newcommand{\be}{\begin{equation}}
\newcommand{\ee}{\end{equation}}
\newcommand{\ben}{\begin{displaymath}}
\newcommand{\een}{\end{displaymath}}
\newcommand{\bea}{\begin{eqnarray}}
\newcommand{\eea}{\end{eqnarray}}
\newcommand{\nn}{\nonumber}

\newcommand{\bean}{\begin{eqnarray*}}
\newcommand{\eean}{\end{eqnarray*}}
\newcommand{\one}{{\rm 1\kern -.9mm l}}

\makeatletter \@addtoreset{equation}{section} \makeatother

\begin{document}

\begin{titlepage}

    \thispagestyle{empty}
    \begin{flushright}
         \hfill{DFPD-09/TH/23} \\
         \hfill{CERN-PH-TH/2009-216} \\
         \hfill{SU-ITP-09/49}\\
         \hfill{ROM2F/2009/23}\\
         \hfill{ENSL-00432620}\\
    \end{flushright}

    \vspace{10pt}
    \begin{center}
        {\LARGE{ \bf A special road to AdS vacua}}

        \vspace{18pt}

        {\bf { Davide Cassani$^{\sharp}$, Sergio Ferrara$^{\diamondsuit,\spadesuit,\flat}$, Alessio Marrani$^{\heartsuit}$,\\ [1mm]
        Jose F.\ Morales$^{\clubsuit}$, and Henning Samtleben$^{\natural}$}}

        \vspace{15pt}
{\small
        {$^{\sharp}$ \it
   Dipartimento di Fisica ``Galileo Galilei"\\
     Universit\`a di Padova, via Marzolo 8, I-35131 Padova, Italy\\
       \texttt{davide.cassani@pd.infn.it}}
       \vspace{5pt}

       {$\diamondsuit$ \it Theory Division - CERN, CH-1211, Geneva 23, Switzerland\\
       \texttt{sergio.ferrara@cern.ch}}
       \vspace{5pt}

        {$\spadesuit$ \it INFN - LNF,
         via Enrico Fermi 40, I-00044 Frascati, Italy}
         \vspace{5pt}

         {$\flat$ \it Department of Physics and Astronomy,\\
       University of California, Los Angeles, CA USA}

        \vspace{5pt}

        {$\heartsuit$ \it Stanford Institute for Theoretical Physics,\\
        Stanford University, Stanford, CA 94305-4060, USA\\
        \texttt{marrani@lnf.infn.it}}
        \vspace{5pt}

        {$\clubsuit$ \it INFN, Universit\`a di Roma ``Tor Vergata'',\\[-.3ex]
via della Ricerca Scientifica, I-00133 Roma, Italy\\[-.3ex]
        \texttt{morales@roma2.infn.it}}
         \vspace{5pt}

         {$\natural$ \it Universit\'e de Lyon, Laboratoire de Physique, ENS Lyon,\\[-.3ex]
         46 all\'ee d'Italie, F-69364 Lyon CEDEX 07, France \\[-.3ex]
         \texttt{henning.samtleben@ens-lyon.fr}}   }
\end{center}

\vspace{5pt}

\begin{abstract}

\noindent We apply the techniques of special K\"ahler geometry to
investigate AdS$_4$ vacua of general ${\cal N}=2$ gauged
supergravities underlying flux compactifications of type~II
theories. We formulate the scalar potential and its extremization conditions in terms of a triplet of prepotentials $\cl P_x$ and their
special K\"ahler covariant derivatives only, in a form that recalls the potential and the attractor equations of ${\cal N}=2$ black holes. We propose a system of first order equations for the $\cl P_x$ which generalize the
supersymmetry conditions and yield non-supersymmetric vacua. Special geometry allows us to recast these equations in algebraic form, and we find an infinite class of new ${\cal N}=0$ and ${\cal N}=1$ AdS$_4$ solutions, displaying a rich pattern of non-trivial charges associated with NSNS and RR fluxes. Finally, by explicit evaluation of the entropy function on the solutions, we derive a U-duality invariant expression for 
the cosmological constant and the central charges of the dual CFT's.
\end{abstract}

\end{titlepage}

\tableofcontents\newpage

\section{Introduction}

The attractor mechanism \cite{Ferrara:1995ih,Strominger:1996kf,Ferrara:1996dd,Ferrara:1996um}, initially discovered in the context of ${\cal N}=2$ black holes, has been recognized as a universal phenomenon governing any extremal flow in supergravity with an AdS horizon. It applies to both BPS and non-BPS solutions \cite{Kallosh:2005ax,Tripathy:2005qp}, ungauged \cite{Sen:2005wa} and gauged \cite{Morales:2006gm} higher-derivatives supergravities, and general intersections of brane solutions \cite{Ferrara:2008xz}.

Also AdS$_4$ vacua arising in string theory compactifications with fluxes can be thought as the near-horizon geometries of extremal black brane solutions. In fact, the vacuum conditions of flux compactifications display many analogies with the attractor equations defining the near-horizon limit of extremal black holes. This was first shown in \cite{Kallosh:2005ax} in the context of  ${\cal N}=1$ orientifolds of type IIB on Calabi-Yau three-folds. In that paper, first it was observed that the ${\cal N}=1$ scalar potential has a form close to the ${\cal N}=2$ black hole potential, with the black hole central charge $Z$ replaced by the combination $e^{K\over 2} W$ involving the ${\cal N}=1$ K\"ahler potential $K$ and superpotential $W$, and with the NSNS and RR fluxes playing the role of the black hole electric charges. Then, exploiting the underlying special K\"ahler geometry of the Calabi-Yau complex structure moduli space, the differential equations both for black hole attractors and for flux vacua were recast in a set of algebraic equations. This treatement did not consider the Calabi-Yau K\"ahler moduli, which do not appear in the type IIB flux superpotential. Explicit attracting Calabi-Yau solutions were derived in \cite{Giryavets:2005nf} (see also \cite{Larsen} for recent developements on CY attractors), while extensions to non-Calabi-Yau attractors were elaborated in \cite{Dall'Agata:2006nr, Anguelova, Kimura:2008tq}. A recent review on the subject, including a complete list of references, can be found in \cite{Bellucci:2007ds}, while for exhaustive reviews on flux compactifications with a thorough bibliography see \cite{FluxReviews}.

In this paper we develop the above ideas focussing on a four-dimensional $\cl N=2$ setup. We consider general ${\cal N}=2$ gauged supergravities related to type II theories via flux compactification on non-Calabi-Yau manifolds, and we  explore their AdS vacua exhibiting partial (${\cal N}=1$) or total ($\cl N=0$) spontaneous supersymmetry breaking. In particular, we find infinite classes of AdS vacua for an arbitrary number of vector and hypermultiplets, and we study their U-duality invariants. The solutions are derived by solving a general system of first order conditions guaranteeing the equations of motion.

At the four dimensional level, the family of supergravities we consider can be obtained by deforming the Calabi-Yau effective action. The latter is characterized by the choice of two special K\"ahler manifolds $\mathscr M_1$ and $\mathscr M_2$: while $\mathscr M_2$ defines the vector multiplet scalar manifold, $\mathscr M_1$ determines via c-map the quaternionic manifold $\mathscr M_Q$ parameterized by the scalars in the hypermultiplets \cite{Cecotti:1988qn,Ferrara:1989ik}. The deformation we study is the most general abelian gauging of the Heisenberg algebra of axionic isometries which is always admitted by $\mathscr M_Q$ \cite{deWit:1992wf, D'Auria:2004tr,D'Auria:2007ay}. The gauging involves both electric and magnetic charges in a consistent way, the latter appearing as mass terms for tensor fields \cite{Louis:2002ny, Theis:2003jj, N=2withTensor1, N=2withTensor2, DeWitSamtlebenTrigiante}. From the point of view of compactifications, this deformation is expected to correspond to a dimensional reduction of type IIA/IIB on $SU(3)$ and $SU(3)\times SU(3)$ structure manifolds (and possibly non-geometric backgrounds) in the presence of general NSNS and RR fluxes \cite{Gurrieri:2002wz, Gurrieri:2002iw, Tomasiello:2005bp, GLW1, KashaniPoor:2006si, GLW2, Cassani:2007pq, Cassani:2008rb}. In this perspective, our setup provides a unifying framework for the study of $\cl N=2$ flux compactifications of type II theories on generalized geometries.

We start our analysis in section~\ref{RevisitingPotential} by reconsidering the expression derived in \cite{D'Auria:2007ay} for the $\cl N=2$ scalar potential associated with the gauging described above. This expression is manifestly invariant under the symplectic transformations rotating the flux charges together with the symplectic sections of both the special K\"ahler geometries on $\mathscr M_1$ and $\mathscr M_2$. We derive a convenient reformulation of the scalar potential in terms of a triplet of Killing prepotentials $\cl P_x\,,\;x=1,2,3$, and their covariant derivatives. The $\cl P_x$ encode the informations about the gauging, and play a role analogous to the central charge $Z$ of $\cl N=2$ black holes, or to the covariantly holomorphic superpotential $e^{\frac{K}{2}}W$ of $\cl N=1$ compactifications. Interestingly, in our reformulation the only derivatives appearing are the special K\"ahler covariant derivatives with respect to the coordinates on $\mathscr M_1$ and $\mathscr M_2$, with no explicit contributions from the $\mathscr M_Q$ coordinates orthogonal to $\mathscr M_1$. This rewriting is advantageous since it allows to take into account the hypersector continuing however to dispose of the nice properties of special K\"ahler geometry.

In section \ref{VacEqs1stOrdCond} we move to the study of the extremization equations for the potential. These are spelled out in subsection~\ref{AttrEqs}, and resemble the attractor equations of ${\cal N}=2$ black holes, though they are more complicated to solve. To achieve this goal, in subsection~\ref{1stOrderCond} we propose a new general approach: this is based on a deformation of the supersymmetry conditions to a set of first order equations in the $\cl P_x$ and their special K\"ahler covariant derivatives, which still imply the (second order) equations of motion and hence yield non-supersymmetric vacua. Again exploiting special K\"ahler geometry, our first order conditions can be reformulated in an algebraic form exhibiting manifest symplectic covariance. The proof that these equations extremize the scalar potential is worked out in subsection~\ref{ExtrFromAnsatz}.

In section~\ref{VacuumSols} we find explicit AdS solutions for the wide class of supergravities whose scalar manifolds $\mathscr M_1$ and $\mathscr M_2$ are symmetric with cubic prepotentials. These solutions generalize those derived in \cite{Cassani:2009ck} in the context of dimensional reduction of type IIA on coset spaces with $SU(3)$ structure (see footnote \ref{PreviousHistory} below for the earlier history of the solutions in \cite{Cassani:2009ck}). Indeed, our analysis allows for an arbitrary number of vector multiplets (subsection~\ref{OneHyper}) and hypermultiplets (subsection~\ref{Many-Hyperss}), as well as for a large set of fluxes. 

Subsection~\ref{U-Invariant-Lambda} is dedicated to the study of the U-duality transformation properties of our solutions. We identify and evaluate the relevant U-invariants, and we provide a manifestly U-invariant expression for the AdS cosmological constant.

In subsection \ref{EntrFct} we compute the central charge of the dual three-dimensional CFT's associated to the solutions found. This is done by employing the {\it entropy function} formalism \cite{Sen:2005wa, Sen:2007qy} originally introduced for black hole attractors, and then generalized to the context of general black brane solutions \cite{Ferrara:2008xz}. The entropy function $F$ is defined as the Legendre transform of the higher-dimensional supergravity action with respect to the brane electric charges, evaluated at the near horizon geometry. Since in our description the brane electric fluxes are assumed dualized to magnetic ones, $F$ corresponds just to (minus) the supergravity action in the four-dimensional description. As we will see, the central charge results then proportional to the inverse of the cosmological constant, and hence determined by the same U-duality invariants mentioned above.
  
In subsection~\ref{StringRealizations} we come back to the explicit examples of type IIA dimensional reductions on coset spaces studied in \cite{Cassani:2009ck} (see also \cite{Caviezel:2008ik} for $\cl N=1$ orientifold truncations on the same manifolds). We give more details about the relevant scalar manifolds, and we discuss the models arising from reduction of type IIB on the same manifolds.

In section \ref{disc} we draw our final considerations. Finally, the appendix contains a lot of background material as well as some details of our computations. In appendix~\ref{saltd} firstly we discuss how the most general scalar potential of $\cl N=2$ supergravity with gaugings of quaternionic isometries can be put in a form involving the $\cl P_x$ and their special K\"ahler and quaternionic covariant derivatives (cf. eq.~(\ref{eq:NewFormV2})). Secondly, we prove that for the theory we consider this expression reduces to a formula in which only special K\"ahler covariant derivatives appear. In appendix~\ref{SummaryN=1} we deal with the $\cl N=1$ susy conditions, expressing them both as differential conditions on the $\cl P_x$, and as symplectically covariant algebraic equations \cite{Cassani:2007pq}. Appendix~\ref{ScalPotIIA} illustrates how the gaugings  we consider can arise from dimensional reduction of type II theories on manifolds with $SU(3)$ structure. Appendix~\ref{U-Invariance} details the computations of the U-duality invariants that are relevant for our solutions.


\section{Revisiting the $\cl N=2$ flux potential}\label{RevisitingPotential}

In this section we reconsider the scalar potential $V$ derived in \cite{D'Auria:2007ay} by gauging the \hbox{$\cl N=2$} effective action for type II theories on Calabi-Yau three-folds, and we reformulate it in terms of Killing prepotentials $\cl P_x$ and their special K\"ahler covariant derivatives.
From a dimensional reduction perspective, $V$ arises \cite{Cassani:2008rb} in general type II flux compactifications on 6d manifolds with $SU(3)$ or $SU(3)\times SU(3)$ structure (and non-geometric backgrounds) preserving eight supercharges \cite{Gurrieri:2002wz, GLW1, KashaniPoor:2006si, GLW2, Cassani:2007pq}. An explicit dictionary between the gauging and the fluxes is presented in appendix \ref{ScalPotIIA} for type IIA on $SU(3)$ structures, together with a further list of references. 

Scalars in ${\cal N}=2$  supergravity organize in vector multiplets and hypermultiplets, respectively parameterizing a special K\"ahler manifold  ${\mathscr M}_{2}$ and a quaternionic manifold ${\mathscr M}_Q$. In the cases of our interest, ${\mathscr M}_Q$ is derived via c-map from a special K\"ahler submanifold ${\mathscr M}_{1}$ \cite{Cecotti:1988qn, Ferrara:1989ik}.
For both type IIA/IIB compactifications, we denote by $h_1+1$ the number of  hypermultiplets (where the 1 is associated with the universal hypermultiplet), and by $h_2+1$ the number of vector fields (including the graviphoton in the gravitational multiplet), so that $h_1={\rm dim}_\C \mathscr M_1 $ and $h_2= {\rm dim}_\C \mathscr M_2$. Furthermore, we introduce complex coordinates $z^i\,,\;i=1,\ldots h_1$ on $\mathscr M_1$, and $x^a\,,\;a=1,\ldots,h_2$ on $\mathscr M_2$, and we denote by
\bea\label{eq:SymplSections}
    \Pi_1^{\bb I} &=&
e^{\frac{K_1}{2}}\left(
\begin{array}{c}
    Z^I \\
    \cl G_I  \\
\end{array}
\right),
\qquad
 I=(0,i)=0,1,\ldots, h_1\,,   \nn\\ [1mm]
   \Pi_2^{\bb A} &=&
e^{\frac{K_2}{2}}\left(
\begin{array}{c}
    X^A \\
    \cl F_A \\
\end{array}
\right),
\qquad A=(0,a)=0,1,\ldots, h_2
\eea
the covariantly holomorphic symplectic sections of the special K\"ahler geometry on ${\mathscr M}_{1}$ and ${\mathscr M}_{2}$ respectively. All along the paper, indices in a double font like $\bb I$ and $\bb A$ correspond to symplectic indices. The respective ranges are $\bb I =1,\ldots, 2(h_1+1)$ and $\bb A=1,\ldots,2(h_2+1)\,$. 

To complete the geometric data on $\mathscr M_{1}$ and $\mathscr M_2$, we introduce the respective K\"ahler potentials $K_1,\,K_2$, K\"ahler metrics $g_{i \bar \jmath},\,g_{a\bar b}\,,$ and symplectic invariant metrics $\C_1,\,\C_2\,$:
\bea
&& K_{1} \, =\,  -\log \,i\big(\,\overline Z^I\cl G_I -  Z^I\overline{\cl G}_I\big) \quad\,
\qquad,\qquad  g_{i\bar \jmath }\,=\,\partial_{i}   \partial_{\bar \jmath} K_1 \nn\\ [1mm]
&& K_{2}  \,=\,  -\log\,i\big(\,\overline X^A\cl F_A -  X^A\overline{\cl F}_A\big)
\qquad  ,\qquad  g_{a \bar b} \, = \, \partial_a   \partial_{\bar b} K_2 \nn \\ [1mm]
\nnb &&
\C_{1\,\bb I \bb J} =\C_1^{\bb I \bb J}=
\left(
\begin{array}{cc}
   \!0 &     \one\!  \\
\!- \one   &    0\!   \\
   \end{array}
\right) = \C_{2\,\bb A \bb B}  =\C_2^{\bb A \bb B}\;\quad\Rightarrow\; \quad \bb C_{\bb{ AB}}\bb C^{\bb{BC}} = -\delta_{\bb A}^{\bb C}\;\;,\;\;\bb C_{\bb{ IJ}}\bb C^{\bb{JK}} = -\delta_{\bb I}^{\bb K}.
\eea
Finally, the coordinates $z^i$ on $\mathscr M_1$ are completed to coordinates on ${\mathscr M}_{Q}$ by the real axions $\xi^{\bb I}=(\xi^I,\tilde\xi_I)^T$ arising from the expansion of the higher dimensional RR potentials, together with the 4d dilaton $\varphi$ and the axion $a$ dual to the NSNS 2--form along the 4d spacetime.

The quaternionic manifold $\mathscr M_Q$ always admits a Heisenberg algebra of axionic isometries \cite{deWit:1992wf}, which are gauged once fluxes are turned on in the higher-dimensional background \cite{D'Auria:2004tr, D'Auria:2007ay, Polchinski:1995sm, Michelson, Dall'Agata:2001zh, Mayr:2000hh, Curio:2000sc}. The NSNS fluxes, including geometric (and possibly non-geometric) fluxes, are encoded in a real `bisymplectic' matrix $Q_{ \bb A}{}^{\bb I}$, while the RR fluxes are encoded in a real symplectic vector $c^{\bb A}$. Explicitly,\footnote{A consistent formulation of $\cl N=2$ supergravity involving both electric and magnetic charges can be obtained by dualizing some hyperscalars to two-forms \cite{Louis:2002ny, Theis:2003jj, N=2withTensor1, N=2withTensor2, DeWitSamtlebenTrigiante}. After a subset of the hypermultiplets is transformed in tensor multiplets, the residual scalar manifold is no more quaternionic. Anyway, it turns out that the correct expression for the scalar potential can be derived by employing the data (vielbeine, $Sp(1)$ connection) of the quaternionic manifold prior the dualization of the scalars, and by reasoning as if the gauging was performed both with respect to electric and magnetic gauge potentials. The resulting expression for $V$ reads as the symplectic completion of the standard formula \cite{Andrianopoli:1996cm} for the potential following from a purely electric gauging. See appendix \ref{saltd} for more details.\\
\indent Also notice that in $\mathscr M_Q$ the special K\"ahler coordinates are inert
under the Heisenberg algebra symmetries of $\mathscr M_Q$ \cite{Ferrara:1989ik, deWit:1992wf}. It is only the latter that is gauged in the supergravity description considered in this paper.
\label{PrecisationMagneticGaugings}}
  \bea\label{DefCharges}
   Q_{\bb A}{}^{\bb I}
&=&
\left(
\begin{array}{cc}
  e_{\!A}^{\;\;I} &   e_{A I}    \\ [1mm]
   m^{A I} &   m^A_{\;\; I}    \\
   \end{array}
\right),
\qquad
\qquad\qquad
c^{\bb A}= \left(
\begin{array}{c}
   p^A   \\ [1mm]
    q_A   \\
   \end{array}
\right)\,.
 \eea
Abelianity and consistency of the gauging impose the following constraints on the charges \cite{D'Auria:2004tr, D'Auria:2007ay}
     \bea \label{const}
     Q^T \C_2 Q = Q\,\C_1 Q^T = c^T Q=0\,.
     \eea
In the context of dimensional reductions, these can be traced back to the Bianchi identities for the higher-dimensional field strengths, together with the nilpotency of the exterior derivative on the compact manifold \cite{GLW1, GLW2} (cf.\ appendix \ref{ScalPotIIA}).

The scalar potential generated by the gauging can be written as the sum of two contributions: $V=V_{\rm NS}+V_{\rm R}$, where $V_{\rm NS}$ can be seen to come from the reduction of the NSNS sector of type II theories, while $V_{\rm R}$ derives from the RR sector. Both these contributions take a symplectically invariant form, and read \cite{D'Auria:2007ay}
\bea
\!\!\!\!\!\!\!\!V_{\rm NS}\! &=& \! -2 e^{ 2\varphi} \Big[ \overline \Pi{}_1^{\,T}  \widetilde Q^T \mathbb{M}_{2} \widetilde Q \Pi_1 + \overline \Pi{}_2^{\,T} Q  \mathbb{M}_{1}   Q^T\Pi_2  +4 \overline \Pi{}_1^{\,T} \C_1^T  Q^T\big( \Pi_2 \overline \Pi{}_2^{\,T} + \overline \Pi_2 \Pi{}_2^T \big) Q\C_1  \Pi_1 \Big]\nn\\ [2mm]
\label{pot0} \!\!\!\!\! V_{\rm R}\! &=&\! -\ft12\, e^{4\varphi} (c+ \widetilde Q  \, \xi  )^T   \mathbb{M}_{2} \,(c+\widetilde Q  \, \xi )\,,
\eea
with
\be\label{DefTildeQ}
\widetilde Q^{\bb A}{}_{\bb I} \,=\, (\C_2^{T} Q \, \C_1)^{\bb A}{}_{\bb I}\,,
\ee
while $(\bb M_1)_{\bb I \bb J}$ and $(\bb M_2)_{\bb A \bb B}$ are symmetric, negative-definite matrices built respectively from the period matrices $(\cl N_1)_{IJ}$ and $(\cl N_2)_{AB}$ of the special K\"ahler geometries on $\mathscr M_1$ and $\mathscr M_2$ via the relation
\be\label{eq:DefbbM}
\bb M\;=\;\left(\begin{array}{cc} 1 & -\re\mathcal N \\ [1mm] 0 & 1 \end{array}\right)
\left(\begin{array}{cc} \im \mathcal N & 0 \\ [1mm] 0 & (\im \mathcal N)^{-1} \end{array}\right)
\left(\begin{array}{cc} 1 & 0 \\ [1mm] -\re\mathcal N & 1 \end{array}\right).
\ee

Nicely, expressions (\ref{pot0}) for $V_{\rm NS}$ and $V_{\rm R}$ can be recast in a form that reminds that of the ${\cal N}=2$ black hole potential (as well as the $\cl N=1$ supergravity potential), with the NSNS and RR fluxes $Q_{ \bb A}{}^{\bb I}$ and $c^{\bb A}$ playing a role analogous to the black hole charges. The black hole central charge will here be replaced by the triplet of $\cl N=2$ Killing prepotentials $\cl P_x\,,\;x=1,2,3$ which describe the gauging under study. These read (see appendix \ref{saltd} for details):
\bea
\nnb \cl P_+ \;\equiv\;  \cl P_1 + i\cl P_2 &=& 2e^{\varphi}\,\Pi_2^T Q \, \C_1 \Pi_1\\ [2mm]
\nnb \cl P_- \;\equiv\; \cl P_1 - i\cl P_2 &=& 2e^{\varphi}\,\Pi_2^T Q \, \C_1 \overline \Pi_1\\ [2mm]
\label{eq:KillingPrep} \cl P_3 &=& e^{2\varphi}\,\Pi_2^T \C_2 (c+\widetilde Q\xi )\,.
\eea
Here, $\cl P_\pm$ encode the contribution of the NSNS sector, while $\cl P_3$ describes the contribution of the RR sector \cite{GLW1, GLW2}.

We find that the NSNS and RR potentials (\ref{pot0}) can be recast in the suggestive form
\bea
V_{\rm NS} &=&
g^{a\bar b}D_a  \cl P_+   D_{\bar b}\overline{\cl P}{}_+ +g^{i\bar \jmath}D_i  \cl P_+  D_{\bar \jmath}\overline{\cl P}{}_+ -2 |\cl P_+|^2\nn\\ [3mm]
V_{\rm R} &=& g^{a\bar b}D_a\cl P_3 D_{\bar b}\overline{\cl P}{}_3 + |\cl P_3|^2, \label{potid}
\eea
whose main benefit is to involve only special K\"ahler covariant derivatives of the $\cl P_x$, which are defined as
\bea
D_i \cl P_x &=& (\partial_i+\ft12 \partial_i K_{1} )\cl P_x \qquad,\qquad D_a \cl P_x = (\partial_a+\ft12 \partial_a K_{2} )\cl P_x\,.
\eea
Eqs.\ (\ref{potid}) are the expressions for $V_{\rm NS}$ and $V_{\rm R}$ we are going to employ in the next sections. An expression closely related to the above rewriting of $V_{\rm NS}$ in terms of $\cl P_+$ appeared in \cite{D'Auria:2007ay}, and our derivation of $V_{\rm R}$ in terms of $\cl P_3$ follows the same methods. In order to prove the equivalence between (\ref{pot0}) and (\ref{potid}) we employ the following useful identities of special K\"ahler geometry \cite{Andrianopoli:2006ub}:
\bea
g^{i\bar \jmath}  D_{i} \Pi_1  D_{\bar \jmath} \overline \Pi{}_1^{\,T}  &=& - \ft12\,   \big(\C_1^T
\mathbb{M}_{1} \C_1 + i\C_1\big) - \overline \Pi_1  \Pi_1^T \label{skind1} \\ [2mm]
g^{a\bar b}   D_{a} \Pi_2  D_{\bar b} \overline \Pi{}_2^{\,T} & =& -  \ft12\, \big( \C_2^T
\mathbb{M}_{2} \C_2+ i\C_2\big) - \overline \Pi_2 \Pi_2^T.\label{skind2}
\eea
These yield
\bea
g^{a\bar b}D_a  \cl P_+ D_{\bar b}\overline{\cl P}_+    &=&   -2\, e^{2\varphi }\big(\,
      \overline \Pi{}_1^{\,T} \widetilde Q^T  \mathbb{M}_{2} \widetilde Q \Pi_1\, +\, 2\overline \Pi{}_1^{\,T}\C_1^T  Q^T   \Pi_2\, \overline  \Pi{}_2^{\,T}\, Q\, \C_1 \Pi_1 \big)\nn\\ [2mm]
     g^{i\bar \jmath}    D_{i}  \cl P_+   D_{\bar \jmath} \overline{\cl P}_+
        &=&   -2 \,e^{2\varphi}\big(\, \overline \Pi{}_2^{\,T}  Q \, \mathbb{M}_{1} Q^T \Pi_2 \,+\,
      2\overline \Pi{}_1^{\,T}\C_1^T  Q^T \Pi_2\, \overline \Pi{}_2^{\,T} Q\, \C_1 \Pi_1\big)\nn\\ [2mm]
      -2 |\cl P_+|^2 &=& - 8\, e^{2\varphi}\, \overline \Pi{}_1^{\,T} \C_1^T  Q^T  \,\overline \Pi{}_2\,  \Pi{}_2^T Q\, \C_1  \Pi_1\nn\\ [2mm]
g^{a\bar b} D_{a} \cl P_3 D_{\bar b} \overline{\cl P}_3 + |\cl P_3|^2  &=&   -\ft12 \,e^{4\varphi}(c + \widetilde Q\xi)^T  \mathbb{M}_{2} (c+ \widetilde Q\xi)\,,\label{eq:ComputationsForV}
\eea
and the equivalence between (\ref{pot0}) and (\ref{potid}) is seen by addition of these four lines.

In appendix \ref{saltd} we illustrate an alternative, {\it ab initio} derivation, where (\ref{potid}) are obtained starting from the general formula for the supergravity scalar potential given as a sum of squares of fermionic shifts, and expressing the latter in terms of the $\cl P_x$ and their derivatives.

It is instructive to compare the quantities in (\ref{eq:KillingPrep}) with the black hole central charge, which reads $Z_{\rm BH}= \Pi_2^T \bb C_2\,c$, where here $c= (p^A,q_A)$ is to be interpreted as the symplectic vector of electric and magnetic charges of the black hole. In addition to the fact that we are dealing with two quantities ($\cl P_+$ and $\cl P_3$) instead of a single one ($Z_{\rm BH}$), we face here the further complication that these do not depend just on the covariantly holomorphic symplectic section $\Pi_2$
of the vector multiplet special K\"ahler manifold $\mathscr M_2$, but also on the scalars in the hypermultiplets. However, the constrained structure of the quaternionic manifold, determined via c-map from the special K\"ahler submanifold $\mathscr M_1$, comes to the rescue, yielding a relatively simple dependence of $\cl P_+$ and $\cl P_3$ on the 4d dilaton $e^\varphi$ (appearing as a multiplicative factor) and on 
the axionic variables $\xi^{\bb I}$ (appearing  in $\cl P_3$ as a scalar-dependent shift of the charge vector $c$). Finally, the $\mathscr M_1$ coordinates enter in $\cl P_+$ via the covariantly holomorphic symplectic section $\Pi_1$, making $\cl P_+$ a covariantly `biholomorphic' object.


\section{Vacuum equations and first order conditions}\label{VacEqs1stOrdCond}

\subsection{The vacuum equations}\label{AttrEqs}

The extremization of the scalar potential (\ref{potid}) corresponds to the equations
\bea
\partial_\varphi V = 0 \,&\Leftrightarrow&\,   V_{\rm NS}+2 V_{\rm R}\,=\,0 \label{eqsatt_phi}\\ [3mm]
\partial_\xi V = 0 \,&\Leftrightarrow& \, \widetilde Q^T \mathbb{M}_{2}\, (c+\widetilde Q\xi)\,=\,0 \label{eqsatt_xi}\\ [3mm]
\label{eqsatt_z} \partial_i V = 0 \, &\Leftrightarrow&\, i C_{ijk} g^{j \bar \jmath}  g^{k \bar k} D_{\bar \jmath}  \cl P_-
   D_{\bar k}  \overline{\cl P}_+\,-\, D_i  \cl P_+  \overline{\cl P}_+ \, + \, g^{a\bar b}D_aD_i \cl P_+  D_{\bar b}\overline{\cl P}_+\, =\,0 \qquad\qquad\\ [3mm]
\nnb \partial_a V = 0\, &\Leftrightarrow &\, iC_{abc} g^{b \bar b} g^{c \bar c}\left( D_{\bar b}  \overline{\cl P}_- D_{\bar c}  \overline{\cl P}{}_+ +  D_{\bar b}  \overline{\cl P}_3  D_{\bar c}  \overline{\cl P}{}_3\right) - D_a  \cl P_+  \overline{\cl P}_+ + 2 D_a  \cl P_3  \overline{\cl P}{}_3 \\ [2mm]
\label{eqsatt_t}  &&   + \,g^{i\bar \jmath}D_i D_a \cl P_+  D_{\bar \jmath}\overline{\cl P}{}_+ \,=\,0\,.
\eea\vskip 1mm
\noindent To write (\ref{eqsatt_z}), we used the following characterizing relations of special K\"ahler geometry \cite{Andrianopoli:1996cm}
\be\label{eq:SKidWithC}
D_iD_j \Pi_1 = i C_{ijk} g^{k\bar k} D_{\bar k} \overline \Pi_1 \qquad,\qquad D_i D_{\bar \jmath} \overline \Pi_1 = g_{i\bar \jmath} \overline \Pi_1\,,
\ee
where $C_{ijk}$ is the completely symmetric, covariantly holomorphic 3--tensor of the special K\"ahler geometry on $\mathscr M_1$. The analogous identities obtained by sending $1\to 2$ and $i,j,k \to a,b,c$ have been used to derive eq.\ (\ref{eqsatt_t}). In particular, these relations imply
\be
D_iD_j \cl P_+ \,=\, i C_{ijk}g^{k\bar k} D_{\bar k} \cl P_-\,,
\ee
as well as
\be
 D_aD_b \cl P_+  \,=\, i C_{abc}g^{c\bar c} D_{\bar c} \overline{\cl P}{}_- \qquad \textrm{and}\qquad D_aD_b  \cl P_3  \,=\, i C_{abc}g^{c\bar c} D_{\bar c} \overline{\cl P}{}_3\,.
\ee
The system of equations above, in particular eqs.\ (\ref{eqsatt_z}) and (\ref{eqsatt_t}), take a form which reminds the attractor equations for black holes in $\cl N=2$ supergravity. 
In the remaining of this section we will show how this set of equations is solved by a supersymmetry inspired set of first
order conditions accounting for both supersymmetric and non-supersymmetric solutions.

\subsection{First order conditions}\label{1stOrderCond}

In this subsection we propose a first order ansatz which generalizes the supersymmetry conditions and allows to solve the vacuum equations (\ref{eqsatt_phi})--(\ref{eqsatt_t}). First we will give a summary of our results, then in the next subsection we will explicitly show how the vacuum equations follow from the ansatz. Finally, in section \ref{VacuumSols} we will present some explicit examples of supersymmetric and non-supersymmetric vacua satisfying our first order ansatz.

We consider the following set of equations, linear in the $\cl P_x$ (and their derivatives), and hence in the flux charges:
\bea
\nnb \pm \,e^{\pm i\gamma-i\theta}\cl P_\pm &=& u\, \cl P_3  \\ [2mm]
 \pm\, e^{\pm i\gamma+i\theta} D_a\cl P_\pm &=& v\, D_a\cl P_3\nn\\ [2mm]
D_i \cl P_+ \;\;= & 0 & =\;\; D_{\bar \imath} \cl P_-\,, \label{eq:1stOrderAnsatz}
\eea
where $\gamma,\,u,\,v,\,\theta$ are real positive parameters. While $\gamma$ will be just a free phase, the other three parameters will need to satisfy certain constraints given below. The ansatz (\ref{eq:1stOrderAnsatz}) generalizes the AdS $\cl N=1$ supersymmetry condition, which, as we illustrate in appendix~\ref{N=1asDP}, corresponds to the particular case\footnote{The $\cl N=1$ conditions are completed by $\cl P_3 = -i\hat{\bar \mu}$, where $\hat \mu\neq 0$ is the parameter appearing in the Killing spinor equation on AdS, related to the AdS cosmological constant $\Lambda$ via $\Lambda = -3|\hat \mu|^2$. See appendix \ref{N=1asDP} for details.\\ For a study of the possible maximally supersymmetric configurations in $\cl N=2$ gauged supergravity, we refer to \cite{Hristov:2009uj}.  \label{Footnotemu}
}
\be\label{eq:SusyValues}
{\textrm{susy}}\quad\Leftrightarrow\quad u=2 \quad , \quad v=1\quad,\quad  \theta=0 \,.
\ee
Our aim is to implement the first order ansatz (\ref{eq:1stOrderAnsatz}) to derive non-supersymmetric solutions of the vacuum equations. 
In order to do this, we will restrict our analysis to the case in which $\mathscr M_2$ is a special K\"ahler manifold with a cubic prepotential. Indeed, below we will show that (\ref{eq:1stOrderAnsatz}) extremizes the scalar potential if the parameters $u,v$ satisfy
\be\label{cond0}
\ft12 u v^2-u^2 v+u+v \,=\, 0\,,
\ee
and --- under the assumption that $\mathscr M_2$ is cubic --- if we further require that
\be
D_a {\cal P}_3 \,=\, \alpha_3\, \partial_a K_2 \, {\cal P}_3 \,,\label{cond2}
\ee
with $\alpha_3$ given by
\bea
e^{3 i{\rm Arg} (\alpha_3)+2 i{\rm Arg} (\cl P_3)}  \,=\, -\sqrt{4u\over 3v} \,
 { 1- v^2\, e^{2i\theta}\over  2- u\,  v\, e^{-2 i \theta} } \;\;,\qquad\quad 
 |\alpha_3|^2 = \frac{u}{3v}\,.  \label{cond3}
 \eea
As we will see, the second of (\ref{cond3}) is actually a consequence of (\ref{eq:1stOrderAnsatz}) and (\ref{cond2}).
Furthermore, notice that by evaluating the modulus square of both its sides, the first of (\ref{cond3}) yields a constraint involving $u,v,\theta\,$ only:
\be
4 \,u \,v^2\cos(2\theta) \,=\, 3\, u^2 v^3 - 4\, u\, v^4 -4\, u+12\, v \,.\label{cond1}
\ee

Explicit solutions to the above conditions on {\it symmetric} scalar manifolds $\mathscr M_2$ with a cubic prepotential will be presented in section \ref{VacuumSols}. 

Below in this section we also consider the possibility of relaxing requirement (\ref{cond2}). While it will be clear that condition (\ref{cond0}) has to hold independently of the assumptions on the special geometry on $\mathscr M_2$, we will show that relation (\ref{cond1}) is also necessary, at least when $\mathscr M_2$ is a {\it symmetric} special K\"ahler manifold.

\vskip 3mm
It will be very useful for our purposes to rephrase eqs.\ (\ref{eq:1stOrderAnsatz}) in a symplectically covariant algebraic form. Following similar steps to the ones presented for the supersymmetric case in appendix \ref{SymplCovForm}, we find that (\ref{eq:1stOrderAnsatz}) are equivalent to\footnote{The equations in appendix \ref{SymplCovForm} are recovered by substituting the values (\ref{eq:SusyValues}) of $u,\, v,\, \theta$, and setting $i\cl P_3=\hat{\bar\mu}$. As shown in \cite{Cassani:2007pq}, these supersymmetry conditions match the `pure spinor equations' derived in \cite{GMPT12, GMPT3} and characterizing the $\cl N=1$ backgrounds at the 10d level.}
\bea
\label{PSE1modif}  0 \!&=&\! Q^T \Pi_2  - i u \cl P_3 e^{ i\theta- \varphi} \re\big(e^{i\gamma}\Pi_1\big) \,,\\ [2mm]
\label{PSE2.1modif} 0 \!&=&\! Q\C_1\re\big(e^{i\gamma} \Pi_1 \big) \,, \\ [2mm]
\nn 0\! &=& \! 2Q \C_1 \im \big( e^{i\gamma} \Pi_1 \big) +2e^{-\varphi}( u + v)
\re\big[ e^{-i \theta} \,\overline{\cl P}_3 \C_2\Pi_2 \big]   -e^{\varphi}(\bb M_2v \cos\theta -\bb C_2 v\sin\theta ) (c+ \widetilde Q \xi).\nn\\
\label{PSE2.2modif}
\eea
Notice that, precisely as in the supersymmetric case, for non-vanishing $\cl P_3$ the second equation actually follows from the first one (cf.\ below eq.\ (\ref{eq:4.36CasBil})).

\subsection{Extremization of $V$ from the Ansatz}\label{ExtrFromAnsatz}

We now come to the proof that the extremization equations (\ref{eqsatt_phi})--(\ref{eqsatt_t}) for the scalar potential $V$ are satisfied by the conditions given above.

\subsubsection*{\underline{$\partial_{\varphi}V=0$}}

We start by considering the extremization of $V$ with respect to the 4d dilaton $\varphi$, namely eq.\ (\ref{eqsatt_phi}). Let us preliminarily show that the linear ansatz above implies
\be\label{eq:NormDPandP}
|D \cl P_3|^2   =\frac{u}{v} |\cl P_3|^2,
\ee
where here and in the following we denote $|D\cl P_x|^2 \equiv g^{a\bar b} D_a \cl P_x D_{\bar b} \overline{\cl P}_x$.
This can be seen from\footnote{An analogous computation leads to $|D\cl P_+|^2 = {\textstyle{\frac{u}{v}}} |v|^2 |\cl P_3|^2$, which is consistent with (\ref{eq:1stOrderAnsatz}) and (\ref{eq:NormDPandP}).}
\bea
\nnb  |D \cl P_3|^2  &=& -\ft 12e^{4\varphi}    \,  (c+\widetilde Q\xi )^T \bb M_2  (c+\widetilde Q\xi ) - |\cl P_3|^2\\ [2mm]
\nnb &=& \frac{( u + v)}{v \cos \theta}e^{2\varphi} (c+\widetilde Q\xi )^T\bb C_2^T  \re\big[ \Pi_2 \overline{\cl P}_3 \,e^{-i\theta} \big] - |\cl P_3|^2\\ [2mm]
&=& \frac{u}{v}\, |\cl P_3|^2,
\eea
where for the first equality we used (\ref{eq:ComputationsForV}), for the second equality we employed condition (\ref{PSE2.2modif}) and the constraints (\ref{const}), while for the third one we recognized expression  (\ref{eq:KillingPrep}) for~$\cl P_3\,$.
From (\ref{eq:1stOrderAnsatz}) and (\ref{eq:NormDPandP}) we deduce the following chain of relations
\be\label{eq:NormsP}
\frac{1}{v} |D\cl P_\pm|^2 \,=\, v\,|D\cl P_3|^2 \,=\, u \, |\cl P_3|^2 \,=\, \frac{1}{u} |\cl P_\pm|^2\,.\\ [2mm]
\ee
Also recalling $D_i \cl P_+=0$, the dilaton equation (\ref{eqsatt_phi}) becomes just a condition on $u$ and $v$:
\bea
0&=& |D{\cl P}_+|^2-2 |{\cl P}_+|^2 + 2(\,|D{\cl P}_3|^2+ |{\cl P}_3|^2\,)\nn\\ [2mm]
&=& \big( u\,v-2 u^2+2{u\over v}+2 \big) |{\cl P}_3|^2,
\eea
which corresponds to eq.\ (\ref{cond0}).

As an aside, we remark that using  (\ref{eq:NormDPandP}) the potential at the critical point can be rewritten as
\be
V=-V_R=-\left(1+\frac{u}{v}\right)|\cl P_3|^2\,. \label{vasp3}
\ee
Notice that in the supersymmetric case this yields $V = -3|\cl P_3|^2 = \Lambda$ (cf.\ footnote~\ref{Footnotemu} for the relation between $\cl P_3$ and the AdS cosmological constant $\Lambda$), which is consistent with the Einstein equation on AdS$_4$.

\subsubsection*{\underline{$\partial_{\xi^{\bb I}} V=0$}}

Next we observe that the $\xi^{\bb I}$ equation (\ref{eqsatt_xi}) follows from eq.\ (\ref{PSE2.2modif}) after multiplying it  from the left  by the matrix $Q^T\bb C_2$. The first and the last terms in the obtained equation vanish due to constraint (\ref{const}), while the second one vanishes after using eq.\ (\ref{PSE1modif}).\footnote{Here we followed parallel steps to the proof \cite[app.~A]{GMPT3} that in type II theories the general $\cl N=1$ conditions of \cite{GMPT3} imply the equations of motion for the RR field-strengths on the internal manifold $M_6$.}
Notice that to satisfy this equation it is crucial that $u$ and $v$ be real.

\subsubsection*{\underline{$\partial_i V=0$}}

Taking into account $D_i \cl P_+ =0$, the only non-trivially vanishing term in (\ref{eqsatt_z}) is the last one, containing both the $D_a$ and $D_i$ derivatives. In order to see that also this term is zero, we evaluate
\bea
\nnb g^{a\bar b}D_aD_i \cl P_+  D_{\bar b}\overline{\cl P}{}_+ &=&
4e^{2\varphi} D_i\Pi_1^T\C_1^TQ^T \left(-\ft 12\C_2^T\bb M_2 \C_2 - \ft i2\C_2 - \overline\Pi_2 \Pi_2^T \right) Q\C_1\overline \Pi_1\\ [2mm]
\nnb &=&  ie^{2\varphi + i\gamma}D_i\Pi_1^T\C_1^T Q^T \Big\{\,2( u + v)  \re\big[ \,\ol{\cl P}_3 e^{-i \theta-\varphi}\bb C_2^T \bb M_2 \Pi_2 \big] \\ [2mm]
\nnb &&\qquad \qquad \qquad \quad+ \; e^{\varphi}  v\, \cos\theta (c + \widetilde Q \xi)  + e^\varphi v\, \sin\theta \,\bb C_2^T \bb M_2(c+\widetilde Q \xi) \Big\} \\ [2mm]
\nnb &=&
\ft 12 e^{i\gamma} (u + v) \big[\,e^{-i \theta} \, \ol{\cl P}_3  D_i\cl P_+ - e^{i\theta}\, \cl P_3  D_i \overline{\cl P}_-\big]\; =\; 0\,.
\eea
where in the first line we used (\ref{skind2}) to obtain the right hand side, whose last two terms actually vanish due to constraint (\ref{const}) and to condition $D_i\overline{\cl P}_- = 0$.  The second line follows from substituting $Q\C_1\overline \Pi_1$
from (\ref{PSE2.1modif}), (\ref{PSE2.2modif}) and using the identity $\C_2^T\bb M_2\C_2 \bb M_2 = \one$. The last line is derived employing again (\ref{const}) and the identity $\bb M_2\Pi_2 = -i\C_2 \Pi_2$, recalling that (\ref{eqsatt_xi}) is satisfied, and recognizing the expressions (\ref{eq:KillingPrep}) of $\cl P_\pm$. Both terms in the parenthesis vanish due to (\ref{eq:1stOrderAnsatz}).

\subsubsection*{\underline{$\partial_aV=0$}}

Finally we consider eq.\ (\ref{eqsatt_t}). Let us first show that the term involving both $D_a$ and $D_i$ derivatives vanishes under the ansatz (\ref{eq:1stOrderAnsatz}). We compute
\bea
\nnb g^{i\bar \jmath}D_iD_a \cl P_+  D_{\bar \jmath}\overline{\cl P}{}_+ &=& 4 e^{2\varphi } D_a\Pi_2^T Q \C_1\big(-\ft 12\C_1^T\bb M_1 \C_1 -\ft{i}{2}\C_1 - \overline\Pi_1 \Pi_1^T \,\big)\C_1^TQ^T\overline\Pi_2  \\ [2mm]
\nnb &=& 2i\, u\, e^{-i \theta}\, \ol{\cl P}_3 e^{\varphi} D_a\Pi_2^T Q \big[\re\big( e^{i\gamma}\bb M_1\Pi_1\big) - 2\C_1\overline \Pi_1 \Pi_1^T\C_1\re\big(e^{i\gamma}\Pi_1\big) \big]\\ [2mm]
\nnb &=&  u\, e^{-i \theta}\,\ol{\cl P}_3 e^{\varphi}D_a\Pi_2^T Q\C_1 \big(e^{i\gamma}\Pi_1 + e^{-i\gamma}\overline\Pi_1\big) \\ [2mm]
\nnb &=&\ft 12\, u\, e^{-i \theta}\, \ol{\cl P}_3 \big(e^{i\gamma}D_a\cl P_+ + e^{-i\gamma}D_a\cl P_-\big)\; =\; 0\,.
\eea
In the first line we used identity (\ref{skind1}); the term  proportional to $i\C_1$ cancels due to constraint (\ref{const}). Then the second line
follows using (\ref{PSE1modif}), while for the third line we used $\bb M_1 \Pi_1 = -i\C_1\Pi_1$ as well as $\Pi_1^T \C_1\overline \Pi_1 = i $. Recalling the expression (\ref{eq:KillingPrep}) of $\cl P_\pm$, we thus obtain the last line, which vanishes due to (\ref{eq:1stOrderAnsatz}).

Using (\ref{eq:1stOrderAnsatz}), eq.\ (\ref{eqsatt_t}) then reads
\be\label{Eqt^aAnsatz}
\big(1 - v^2e^{2i\theta} \big) iC_{abc}g^{b\bar b}g^{c\bar c}D_{\bar b}\overline{\cl P}_3 D_{\bar c}\overline{\cl P}_3 \,+\, \big(2-uv e^{-2i\theta}\big)D_a \cl P_3 \,\overline{\cl P}_3 \,=\, 0\,.
\ee
In the supersymmetric case (\ref{eq:SusyValues}), and only in this case, the terms in the two parenthesis vanish separately, and the scalar potential is therefore fully extremized.

To solve the non-supersymmetric vacuum equations we focus on the case in which $\mathscr M_2$ has a cubic prepotential, and require relation (\ref{cond2}). This condition, together with the following property of cubic geometries\footnote{This can be verified using the relations given in (\ref{cds2}), (\ref{eq:RelCubicSKgeo}) below.\label{RefToCubicRel}}
\be
iC_{abc} g^{b\bar b}g^{c\bar c}\,\partial_{\bar b}K_2\, \partial_{\bar c}K_2 \,=\, 2\,\partial_a K_2\,,
\ee
allows us to rewrite (\ref{Eqt^aAnsatz}) as
\be
\big(1 - v^2e^{2i\theta} \big) 2\bar \alpha_3^2\, \overline{\cl P}{}_3^2  \,+\, \big(2-uv e^{-2i\theta}\big) \alpha_3 |\cl P_3|^2    \,=\, 0\,.
\label{eqda}
\ee
where we used the relation $g^{a\bar b}\partial_aK_2\partial_{\bar b}K_2=3$   (cf.\ footnote \ref{RefToCubicRel}), 
that holds for cubic geometries. Using  the second of (\ref{cond3}) it is now easy to check that (\ref{eqda}) corresponds to the first of (\ref{cond3}).
\vskip 4mm
We have thus proved that conditions (\ref{eq:1stOrderAnsatz}), together with (\ref{cond0})--(\ref{cond3}), guarantee (a non-supersymmetric) extremization of the scalar potential.

\newpage
 
\subsubsection*{Relations between symplectic invariants}

One can wonder whether it is possible to find vacuum solutions of the type (\ref{eq:1stOrderAnsatz}) that do not require ansatz (\ref{cond2}). Here we show that, even relaxing the latter condition, eq.\ (\ref{cond1}) is however still needed, at least in the case in which the special K\"ahler manifold $\mathscr M_2$ is {\it symmetric}. 

We start by writing eq.\ (\ref{Eqt^aAnsatz}) as
\bea\label{Eqt^aAnsatz2}
D_a \cl P_3 &=&{A\over   \overline{\cl P}_3 } \, C_{abc}g^{b\bar b}g^{c\bar c}D_{\bar b}\overline{\cl P}_3 D_{\bar c}\overline{\cl P}_3\;,
 \qquad ~~~~~~~~~~~
A\equiv -{ i\big(1 - v^2e^{2i\theta} \big) \over  \big(2-uve^{-2i\theta}\big) }\,.
\eea
Contracting  this equation with $g^{a \bar b} D_{\bar b} \cl P_3$ one finds a necessary condition relating two symplectic invariants:
\be
|DP_3|^2={A\over   \overline{\cl P}_3 }\, C_{abc}g^{a\bar a} g^{b\bar b}g^{c\bar c}\,D_{\bar a}\overline{\cl P}_3 \, D_{\bar b}\overline{\cl P}_3 D_{\bar c}\overline{\cl P}_3\,.   \label{identans}
\ee
Replacing $ D_{\bar b} \overline{\cl P}_3$ and $ D_{\bar c} \overline{\cl P}_3$ in (\ref{Eqt^aAnsatz2}) by the complex conjugate of (\ref{Eqt^aAnsatz2}) itself and using the following identity, valid for symmetric special K\"ahler geometries:
\begin{equation}
g^{b\overline{b}}g^{c\overline{c}}\, C_{a bc}%
C_{\bar b(\bar  d\bar e}C_{\bar  f\bar g)\bar c}=\frac{4}{3}C_{(\bar d\bar e\bar f}g_{\bar  g) a} \,,
\end{equation}
we get
\be
D_a \cl P_3 \, =\, {4\,A\, \bar A^2\over   3\, \overline{\cl P}_3\, \cl P_3^2 }  \,D_{a} \cl P_3\,  C_{\bar b \bar c \bar d}\,  g^{\bar b b}g^{\bar c c} g^{\bar d d} D_{b} \cl P_3 \,D_{c}  \cl P_3 \,D_{d} \cl P_3\,.
\ee
Using (\ref{identans}) and (\ref{eq:NormDPandP}) to rewrite the right hand side, we arrive at
\be
{4\,u\,\over  3\, v \, }  |A|^2 \,=\, 1\,,
\ee
which is precisely eq.\ (\ref{cond1}).

\subsubsection*{Explicit solutions}

In the next section we will present explicit vacuum solutions ${\bf A}$, ${\bf B}$, ${\bf C}$ (cf. (\ref{sola})--(\ref{solc})) satisfying the linear ansatz of subsection \ref{1stOrderCond}, with the parameters $u,v,\theta$ given by
\bea
\nnb  {\bf A}  \;\;(\cl N=0) &:&\quad u =3 v=2\sqrt{\ft65}
  \;,\qquad   e^{i\theta}={\textstyle{\frac{1}{\sqrt{6}} }} (\sqrt{5}- i)  \\ [3mm]
\nnb {\bf B} \;\;(\cl N=0)&:&\quad u \,=\,v\,\,=\,\,2\quad \:, \qquad e^{i\theta}=\ft12(1-i\sqrt 3)  \\ [3mm]
{\bf C}  \;\;(\cl N=1) &:&\quad u\,=\,2 v\,=\, 2   \quad , \qquad  e^{i\theta}=1  \,.
\label{soluvth}
\eea

  \section{Vacuum solutions}\label{VacuumSols}

Applying the results of the previous section, in the following we present new explicit ${\cal N}=0$ and $\cl N=1$ AdS$_4$ vacuum solutions of the ${\cal N}=2$ gauged supergravities under study, and we discuss the associated U-duality invariants. 

For simplicity we focus on the
case where the special K\"ahler scalar manifolds ${\mathscr
M}_{1}$ and ${\mathscr M}_{2}$ are symmetric manifolds $G/H$ with cubic prepotentials; a complete list is given in table \ref{cubicSKandQ} below. We denote the cubic prepotentials on  ${\mathscr M}_{1}$ and ${\mathscr M}_{2}$ respectively by
  \be
  \cl F= \ft16 d_{abc} {X^a X^b X^c\over X^0} \qquad\;\;,\;\; \qquad   \cl G=\ft16 d_{ijk} {Z^i Z^j Z^k\over Z^0} \,,    \qquad
  \ee
where $d_{abc}$ and $d_{ijk}$ are scalar-independent, totally symmetric real tensors.

\begin{table}[!b]
\begin{center}
$\begin{array}{|c|c|c|c|c|c|c|}
\hline
\rule{0pt}{3ex}
{G\over H} &   \frac{SU(1,1)}{U\left( 1\right) } &  \frac{SU(1,1)}{U\left( 1\right) }\times \frac{SO(2,n)}{SO(n)\times U\left(
1\right) }   &     \frac{Sp(6,\mathbb{R})}{SU\left( 3\right) \times U\left( 1\right) }&     \frac{SU(3,3)}{SU\left( 3\right) \times SU\left( 3\right) \times U\left(1\right) }&    \frac{SO^{\ast }(12)}{SU\left( 6\right) \times U\left( 1\right) } &  \frac{E_{7\left( -25\right) }}{E_{6}\times SO\left( 2\right) } \\ [3mm]
{\mathscr M}_Q &    \frac{G_{2\left( 2\right) }}{SO\left( 4\right) }      &\frac{SO(4,n+2)}{SO(n+2)\times SO\left( 4\right) } &
 \frac{F_{4\left( 4\right) }}{USp\left( 6\right) \times SU\left( 2\right) } & \frac{E_{6\left( 2\right) }}{SU(6)\times SU\left( 2\right) } &
\frac{E_{7\left( -5\right) }}{SO(12)\times SU\left( 2\right) } &    \frac{E_{8\left( -24\right) }}{E_{7}\times SU\left( 2\right) }\\ [2mm]
\hline
\rule{0pt}{3ex}
{\bf R}_G &  {\bf 2} & ({\bf 2},{\bf n+2}) & {\bf 14} & {\bf 20} & {\bf 32} & {\bf 56} \\ [1mm]
{\bf R}_H & 0  & {\bf n}+{\bf 1} & {\bf 6} & ({\bf 3},{\bf 3})  & {\bf 15} & {\bf 27} \\ [1mm]
\hline
\end{array}$
\caption{The first row displays the complete list of symmetric special K\"ahler manifolds $G/H$ with cubic prepotentials (see \textit{e.g.} \cite{GunSierraTown, deWit:1992wf}). The second row shows the quaternionic manifolds ${\mathscr M}_{Q}$ related to the  special K\"ahler manifolds in the first row via the c-map \cite{Cecotti:1988qn}. The third row displays the symplectic $G$-representation under which the symplectic sections ($\Pi_1^{\bb I}$ or $\Pi_2^{\bb A}$) transform. Finally, the last row shows the $H$-representation under which the scalar coordinates ($z^i$ or $x^a$) on $G/H$ transform.
}\label{cubicSKandQ}
\end{center}
\end{table}

Choosing special coordinates, for $\mathscr M_2$ we have
  \bea
  X^A \,=\, (1,x^a) \quad\,,\,\quad \cl F_A=(-f,f_a)   \quad &,& \quad f= \ft16 d_{abc} x^a x^b x^c \quad\,,\,\quad  f_a = \ft12 d_{abc} x^b x^c  \nn\\ [2mm]
\nnb {\cal V}\,=\,\ft16 d_{abc} x_2^a x_2^b x_2^c \quad ,\quad \cl V_a = \ft 12 d_{abc} x_2^b x_2^c \quad&,&\quad  C_{abc}= e^{K_{2} } \, d_{abc} \\ [2mm]
K_2=-\log(-8 {\cal V}) \quad &,&\quad\partial_aK_2={i {\cal V}_a\over 2 {\cal V}}\:,\label{cds2}
\eea
where the complex coordinates $x^a$ are split into real and imaginary parts as $x^a = x_1^a +ix_2^a\,$, with $x_2^a<0$.
Analogously, for $\mathscr M_1$ we have
\bea
Z^I \,=\, (1,z^i) \quad\,,\,\quad   {\cal G}_I=(-g, g_i)   \quad &,&\quad  g=\ft16 d_{ijk} z^i z^j z^k \quad\,,\,\quad  g_i=\ft12 d_{ijk} z^j z^k  \nn\\ [2mm]
\nnb \tilde{\cal V}\,=\,\ft16 d_{ijk} z_2^i z_2^j z_2^k \quad , \quad \tilde{\cal V}_i\,=\,\ft12 d_{ijk} z_2^j z_2^k\quad &,&\quad   C_{ijk}= e^{K_{1} } \, d_{ijk} \\ [2mm]
 K_1= -\log(-8 \tilde{\cal V})\quad &,& \quad \partial_iK_1 ={i {\tilde{\cal V}}_i\over 2 {\tilde{\cal V}}}\:.\label{cds1}
\eea
with $z^i=z_1^i+iz_2^i$ and $z_2^i<0$.
For the case of a theory with no hypermultiplets other than the universal one, expressions (\ref{cds1}) are replaced simply by $Z^0=1\,,\,\cl G_0= -i$ and $e^{K_1}=\ft 12$.

Moreover, the following relations involving the metric $g_{a\bar b} = \partial_a\partial_{\bar b}K_2$ on ${\mathscr M}_{2}$ are valid:
\bea
\nnb g_{a \bar b}\, =\,\frac{1}{4{\cal V}^2} ({\cal V}_a{\cal V}_b-{\cal V} \,{\cal V}_{ab} )\quad &,& \quad g^{a \bar b} \,=\, 2 ( \,x_2^a \, x_2^b -2{\cal V} \,{\cal V}^{ab})  \\ [2mm]
\label{eq:RelCubicSKgeo}
g^{ab}{\cal V}_b \,= \,4{\cal V} x_2^a \quad &,&\quad  g^{ab}{\cal V}_a{\cal V}_b \,=\, 12{\cal V}^2,
\eea
where $\cl V_{ab} = d_{abc}x_2^c\,$, and ${\cal V}^{ab}$ is its inverse. Similar relations hold for the corresponding quantities on ${\mathscr M}_{1}$ with $a,b\to i,j$ and ${\cal V}\to \tilde{\cal V}$.

To perform the forthcoming computations, it is convenient to introduce the following holomorphic prepotentials $W_{\pm},\,W_3\,$:
\bea\label{eq:DefW}
{\cl P}_{\pm} &=& e^{{K_1+K_2\over 2}+\varphi} \, W_{\pm} \;\,, \qquad \qquad{\cl P}_{3} =e^{{K_2\over 2}+2\varphi} \, W_{3} \,, 
\eea
 whose K\"ahler covariant derivatives are defined via
 \bea
 D_a W_x &=& (\partial_a +  \partial_a K_2) W_x  \,,\nn\\ [2mm]
 D_i W_{+} &=& (\partial_i + \partial_i K_1) W_{+} \,,\qquad\qquad \partial_i W_-\, =\, 0\,, \nn \\ [2mm]
    D_{\bar \imath} W_{-} &=& (\partial_{\bar\imath} + \partial_{\bar\imath} K_1) W_{-} \,,\qquad\qquad  \partial_{\bar\imath} W_+ \,=\, 0\,,
 \eea
with $\partial_a K_2$ and $\partial_i K_1$ given in (\ref{cds2}) and (\ref{cds1}) respectively.
 Explicitly, recalling (\ref{eq:DefW}), (\ref{eq:KillingPrep}), and taking $m^{AI}=m^A{}_I=0$ for simplicity, one finds
 \bea
 W_+ &=& 2 X^A( e_A{}^I \cl G_I- e_{AI} Z^I) \;\;,\qquad\qquad W_-\;=\;  2X^A( e_A{}^I \bar{\cl G}_I- e_{AI} \bar Z^I)\,, \nn\\ [2mm]
 W_3&=&  X^A  (q_A+e_{A}{}^I \tilde\xi_I-e_{AI}\xi^I) -{\cal F}_A p^A\,,
 \eea
Recalling (\ref{cds2}) and defining $f_{ab}=d_{abc}x^c$, we preliminarily compute:
 \bea
 \partial_a W_+ &=& 2( e_a{}^I \cl G_I-  e_{aI}Z^I) \;\;,\qquad\qquad 
\partial_a  W_-\;=\;  2(e_a{}^I \bar{\cl G}_I- e_{aI}\bar Z^I)\,, \nn\\ [2mm]
 \partial_a W_3&=&   q_a+e_{a}{}^I \tilde\xi_I-e_{aI} \xi^I + f_a p^0-f_{ab} p^b\,.
 \eea
To solve the vacuum equations (\ref{eqsatt_phi})--(\ref{eqsatt_t}) in full generality is a challenging problem that goes beyond the scope of this paper. In the following we present some simple solutions as prototypes of the general case.

\subsection{Only universal hypermultiplet}\label{OneHyper}

We start by considering the case of a gauged supergravity with a single hypermultiplet (i.e. $h_1=0$), which we identify with the universal hypermultiplet of string compactifications. Concerning the vector multiplets, we allow for an arbitrary number of them, and we just require that the associated special K\"ahler scalar manifold is symmetric and has a cubic prepotential,\footnote{In particular, the second property is relevant for type IIA compactifications on 6d manifolds $M_6$ with $SU(3)$ structure, where the K\"ahler potential $K_2$ is expected to take the cubic form $e^{-K_2}= \frac{4}{3}\int_{M_6} J\wedge J\wedge J$, where $J$ is the almost symplectic 2--form on $M_6$. See subsection \ref{StringRealizations} and appendix \ref{ScalPotIIA} for more details.} specified by the 3-tensor $d_{abc}$. As it will be clear in the following, the latter assumption will allow us to perform computations in a more explicit fashion.
In addition we assume that the only non-vanishing entries of the charge matrix $Q$ be $e_{a0}\equiv e_a$. We remark that this choice might be generalized by using U-duality rotations. 
The constraints (\ref{const}) require that $e_a p^a=0$\,.
For this choice of charges the  prepotentials become
 \bea
 W_{\rm +} &=& W_-=  - 2e_a x^a \,, \nn\\ [2mm]
W_3 &=& q_0 + (q_a - e_{a} \xi) x^a  +p^0 f-p^a f_a \,,  \label{prep13}
 \eea
where all along this subsection we denote $\xi \equiv \xi^0$.

It is convenient to introduce the following shifted variables (assuming $p^0 \neq 0$):
\be\label{eq:DefBoldVars}
{\bf x}^a = x^a-{p^a\over p^0} \,,\qquad \qquad{\bf q}_0=q_0+{q_a p^a\over p^0}  -{2 P\over (p^0)^2}   \,,\qquad \qquad   {\bf q}_a=q_a-{P_a\over p^0}\,,
\ee
      with
   \bea
   P &=& \ft16 d_{abc} p^a p^b p^c \,,\qquad  P_a=\ft12 d_{abc}  p^b p^c\,.   \label{shift}
   \eea
   In terms of these variables one finds
    \bea
 W_+ &=&  W_-= - 2e_a {\bf x}^a \,,\qquad\qquad  ~~~~~~~~~~~   \partial_a W_+=  \partial_a W_-= - 2e_a\,,\nn\\ [2mm]
  W_3  &=&    {\bf q}_0 +({\bf q}_a-\xi e_a) {\bf x}^a  +p^0 {\bf f} \,, \qquad
 ~~~~~~\partial_a W_3=  {\bf q}_a-\xi e_a  +p^0 {\bf f}_a  \,, \label{prep132}
 \eea
with
   \be
     {\bf f}=\ft16 d_{abc} {\bf x}^a  {\bf x}^b  {\bf x}^c\;,\qquad\quad  {\bf f}_a=\ft12 d_{abc}    {\bf x}^b  {\bf x}^c\,.
   \ee
In writing (\ref{prep132}) we have used that $e_a p^a=0$.
Notice that since ${\bf x}_2^a \equiv x_2^a$, the expressions in (\ref{eq:RelCubicSKgeo}) can be equivalently written with the bold variable. The advantage of using the bold variables introduced above is that the explicit dependence on $p^a$ is entirely removed.
Finally we introduce the following quantities built from the NSNS fluxes $e_a\,$:
 \be
   R=\ft16 d^{abc} e_a e_b e_c \,,\qquad R^a=\ft12 d^{abc}   e_b e_c \label{R-def}\,,
   \ee
where $d^{abc}$ is the contravariant tensor of the symmetric special K\"ahler geometry satisfying
\be
d_{abc} \,d^{b (d_1 d_2} \,d^{d_3 d_4) c} =\ft43 \, \delta_a^{(d_1} d^{d_2 d_3 d_4)}. \label{idsym}
\ee
Solutions of the vacuum equations can be found starting from the simple ansatz
         \be
     {\bf x}^a =x \, R^a\,,  \qquad  {\bf q}_a =0\,,
     \label{simple}
     \ee
    with $x=x_1+i x_2$ a complex  function of the charges to be determined. This ansatz can be motivated by noticing that once ${\bf q}_a$ is taken to zero, the only contravariant vector one can build with $e_a$'s variables is $R^a$.
   Using (\ref{idsym}) one finds the following relations:
 \bea
    d_{abc} \,R^b R^c =  2 \,R\, e_{a} \,,\qquad &&  R^a \,e_a=3 \,R\,,   \nn\\ [2mm]
     {\cal V} = (x_2)^3 R^2\,, \qquad    {\cal V}_a =(x_2)^2 R\, e_a\,, \qquad &&     {\bf f }=x^3 R^2\,, \qquad     {\bf f}_a =x^2  R\, e_a\,,\nn\\ [2mm]
    W_{\pm} = -6\,R\, x   \,, \qquad &&
  W_{3} = ({\bf q}_0-3R\, \xi\, x  + p^0 R^2  x^3 )\,,   \nn\\ [2mm]
    \partial_a W_+ =  \partial_a W_-= - 2e_a\,,\qquad &&
  \partial_a W_3=  (\,p^0 R\, x^2 -\xi \, ) e_a\,, \nn\\ [2mm]
  e^{-K_1}=2\,, \qquad &&  e^{-K_2}=-8{\cal V}\,.
  \label{eq:Wansatz}
  \eea
Moreover, the covariant derivatives of the prepotentials  take the form
\be \label{eq:DWansatz}
   D_a  W_{x}  \,=\,  \partial_a K_2 \, \alpha_{x}   W_x \,=\, \frac{ie_a}{2x_2R} \, \alpha_{x}   W_x   \,, 
\ee
with     
   \bea \label{alpha3abc}
   \alpha_{\pm} =  1-  {2i x_2  \over 3 x }    \,, \quad \qquad
   \alpha_{3} =   1- {2i R x_2 (\,p^0 R \,x^2 -\xi\,) \over  {\bf q}_0-3\,R\, \xi\, x + p^0 R^2 x^3   } \,.
         \eea
Using the relation (here there is no sum over $x$):
\be
g^{a \bar b} D_a \cl P_x  D_{\bar b} \overline{\cl P}_x \,=\, 3 |\alpha_x \cl P_x|^2\,,
\ee
the scalar potential reads
   \bea
   V &=&e^{K_1+K_2+2\varphi}\,  \big(3 |\alpha_+|^2-2  \big)|W_+|^2\:+\: e^{K_2+4\varphi}\, \big(3 |\alpha_3|^2+1  \big)|W_3|^2
   \,. \label{vfin}
   \eea

Let us now combine the ansatz (\ref{simple}) with the linear ansatz (\ref{eq:1stOrderAnsatz}) that we have established in
the previous section. It is straightforward to verify that with (\ref{eq:Wansatz}), (\ref{eq:DWansatz}), the first 
two equations of (\ref{eq:1stOrderAnsatz}) are identically satisfied upon defining $u, v, \theta, \gamma$ as
\be
\!\!\!u\, e^{i\theta} = 6i e^{K_1/2-\varphi} \,\frac{R\, x}{ {\bf q}_0-3\,R\, \xi\, x+p^0R^2 x^3   }\;,
\quad\;\;\;\;
v~=~\frac{\alpha_+ u\, e^{2i\theta}}{\alpha_3}
\;,
\quad\;\;\;\;
e^{-i\gamma}=i
\;. \label{uvtheta}
\ee
Then the second equation of (\ref{cond3}) can be solved for $\xi$ and yields
\bea
\xi&=& 
\frac{{\bf q}_0 x_1 + p^0 R^2 (x_1^4+x_1^2x_2^2)}{R(3x_1^2+x_2^2)} ~=~
 \left(\frac{p^0 {\bf q}_0^2}{R}\right)^{1/3}\,
 \frac{\chi_1 (1+\chi_1^3+\chi_1\chi_2^2)}{3\chi_1^2+\chi_2^2}\;,
 \label{xi}
\eea
where we have rescaled $x_{1,2}$ as\footnote{
For simplicity, we assume a sector of charges, where ${\bf q}_0, p^0, R > 0$\,.
}
\bea
x=x_1+i x_2 \equiv \left(\frac{{\bf q}_0}{p^0 R^2}\right)^{1/3}\,(\chi_{1}+i\chi_2)\,,\label{xx1x2}
\eea 
in terms of dimensionless quantities $\chi_{1,2}$\,.
This also guarantees that $v$ is real.
Likewise, equation (\ref{cond0}) can be solved for $\varphi$ and yields
\bea
e^{2\varphi} &=& \frac34\, \left(\frac{R}{p^0 {\bf q}_0^2}\right)^{2/3}\,
\frac{(3\chi_1^2+\chi_2^2)(-3\chi_1^2+5\chi_2^2)}{\chi_2^2(1-4\chi_1^3+4\chi_1^6+9\chi_1^4\chi_2^2+6\chi_1^2\chi_2^4+\chi_2^6)}
\;.
\label{phi}
\eea
Moreover, from (\ref{vfin}) we find for the scalar potential
\bea
   V &=&
 -  \frac9{32}\, \left(\frac{R^4}{p^0 {\bf q}_0^5}\right)^{1/3}\,
\frac{(3\chi_1^2+\chi_2^2)(-3\chi_1^2+5\chi_2^2)^2}{\chi_2^5(1-4\chi_1^3+4\chi_1^6+9\chi_1^4\chi_2^2+6\chi_1^2\chi_2^4+\chi_2^6)}
   \,. 
\eea
Remarkably, all dependence on the charges factors out. 

It remains to solve the first equation of (\ref{cond3}), or equivalently (\ref{eqda}). 
Plugging  (\ref{eq:Wansatz})--(\ref{xx1x2}) into (\ref{eqda}), this complex equation
finally gives rise to two real polynomial equations in $\chi_{1,2}$, which can be solved explicitly, and admit precisely three real solutions
     \bea
    {\bf A:}   \qquad &&  
    \chi_1=0\,,\;       \qquad  ~~~~~~~~  \;\chi_2=-5^{-1/6}  \nn\\  
    {\bf B:}    \qquad &&        \chi_1=20^{-1/3}  \,,  \qquad   ~~~~~~~\chi_2 = -\sqrt{3}\,20^{-1/3}  \nn\\  
    {\bf C:} \qquad &&     \chi_1=- \ft12 20^{-1/3} \,, \qquad  ~~~~~~~~ \chi_2 = - \ft12 \sqrt{15}\,20^{-1/3} \label{solc0}
    \eea
  Putting everything together, the three solutions are given by
     \begin{itemize}
    \item{Solution {\bf A} $\,$($\cl N=0$) :
    \bea
   x_1 &=& \xi=0\;,\qquad x_2=-5^{-{1\over 6}}   \left(\frac{{\bf q}_0}{p^0 R^2}\right)^{1/3} \,,  \qquad
        e^\varphi = \ft1{2\sqrt{2}} \,5^{5\over 6}\,   \left( { \ R\over   p^0 {\bf q}_0^2 }\right)^{1/3}\,,  \nn\\ [3mm]
    V &=&   -\ft{75}{64}  \,5^{5\over 6}\,   \left( { \ R^4\over   p^0 {\bf q}_0^5 }\right)^{1/3}\,. \label{sola}
    \eea
    }
       \item{Solution {\bf B} $\,$($\cl N=0$) :
    \bea
   x_1 \!\!&=&\!\!  -\ft{1}{\sqrt{3}} \,  x_2 =     \left(\frac{{\bf q}_0}{20 p^0 R^2}\right)^{1/3},
    \quad \xi=\left( { \ 4 p^0 {\bf q}_0^2 \over  25 R }\right)^{1/3}\,,\quad
    e^\varphi = \ft{\sqrt{2}}{\sqrt{3}} \,   \left( { 25  R\over  4 p^0 {\bf q}_0^2 }\right)^{1/3}, \nn\\ [3mm]
      V  &=&  - \ft{5}{\sqrt{3}} \,      \left( { 25\, R^4\over   4 p^0 {\bf q}_0^5 }\right)^{1/3}\,. \label{solb}
    \eea
    }
        \item{Solution {\bf C} $\,$($\cl N=1$) :
    \bea
    x_1 \!\!&=&\!\!\ft{1}{\sqrt{15}} \,  x_2 = - \ft12   \left(\frac{{\bf q}_0}{20p^0 R^2}\right)^{1/3},
         \quad
     \xi= -\left( { \  p^0 {\bf q}_0^2 \over  50 R }\right)^{1/3}\,,\quad
    e^\varphi = \ft{\sqrt{2}}{\sqrt{3}} \,5^{1\over 6}\,\left( { 2  R\over   p^0 {\bf q}_0^2 }\right)^{1/3},
    \nn\\ [3mm]
    V   &=&  - 8 \sqrt{3} \, 5^{-{5\over 6}}      \left( { 2\, R^4\over   p^0 {\bf q}_0^5 }\right)^{1/3}\,. \label{solc}
    \eea
    }
     \end{itemize}
The corresponding values of $u, v, \theta$ have been given in (\ref{soluvth}) above. Notice that the assumed positivity of the charges guarantees $e^\varphi >0$, $x_2<0$ and $V<0$. 

The above solutions generalize to an arbitrary number of vector multiplets the ones derived in \cite{Cassani:2009ck} in the context of flux compactifications of type IIA on coset manifolds with $SU(3)$ structure.\footnote{See subsection \ref{StringRealizations} for some more details. The $\cl N=1$ solutions appearing in \cite[sect.$\:$6]{Cassani:2009ck} were already known: they were first found at the 10d level in \cite{Behrndt:2004km}, studied from a 4d perspective in \cite{House:2005yc, KashaniPoorNearlyKahler}, and extended in \cite{Tomasiello:2007eq, Koerber:2008rx}. The $\cl N=0$ solutions had already been derived, from a 10d perspective, in \cite{Romans:1985tz, Lust:2008zd}. Further $\cl N=0$ AdS vacua on the same cosets with non-zero orientifold charge where found in \cite{Danielsson:2009ff}. The general conditions for supersymmetric AdS vacua of type IIA on $SU(3)$ structures were first given in \cite{LustTsimpis1}. \label{PreviousHistory}} Furthermore, they allow for non-vanishing charges $p^a,q_a$ (satisfying ${\bf q}_a=0$, namely $p^0 q_a=\ft 12 d_{abc}p^bp^c$).

 \subsection{Adding hypermultiplets}\label{Many-Hyperss}

The three solutions above can be generalized to the case of a cubic supergravity with arbitrary number of vector multiplets and hypermultiplets. For simplicity we focus again to the case where the vector multiplet scalar manifold is symmetric. Here we consider a vacuum configuration with non-trivial charges:
 $e_a{}^i$, $e_a\equiv e_{a0}$, $p^0$, ${\bf q}_0$, while ${\bf q}_a$ and all remaining charges
in $Q_{\bb A}{}^{\bb I}$ are set to zero.
Recalling (\ref{eq:DefBoldVars}), the prepotentials and their derivatives are now given by
\bea
 W_+ &=&  2{\bf x}^a( e_a{}^i g_i- e_{a}) \;,\qquad   \partial_a W_+= 2(e_a{}^i g_i- e_{a}) \; ,\qquad
   \partial_i W_+  = 2{\bf x}^a e_a{}^j g_{ij}\,,\nn\\ [2mm]
W_3&=&  {\bf q}_0+ {\bf x}^a (e_{a}{}^i \tilde\xi_i-e_{a} \xi^0 )  + p^0 {\bf f} \;,\qquad\qquad  \partial_a W_3= e_{a}{}^i \tilde\xi_i-e_{a} \xi^0+ p^0{\bf f}_a  \,,
\eea
where $g_{ij}=d_{ijk}z^k$. The expressions for $W_-$ and its derivatives are like those for $W_+$, with $g_i\to \bar g_i$ and $g_{ij}\to\bar g_{ij}$. Again we follow an educated ansatz for the solution:
  \be
  {\bf x}^a=x\, R^a \,,\qquad ~~~~~~  z^i=z\, S^i \,,\qquad~~~~~ \tilde\xi_i=\zeta \,T_i\,,
  \ee
where $R^a$ is defined as in (\ref{R-def}), and we introduced the combinations of NSNS charges
  \bea
  S^i\,=\, e_a{}^i R^a  \quad &,&\quad T \,=\,\ft16 d_{ijk} S^i S^j S^k \quad\; ,\;\quad T_i=\ft12 d_{ijk}  S^j S^k\,.
  \eea
    In addition we impose the following relation
\be\label{Constraintond}
d_{ijk} e_a{}^i e_b{}^j e_c{}^k =\beta\, d_{abc}      \qquad \Rightarrow  \qquad  T_i e_a{}^i=\beta\, R\, e_a  \quad\;,\;\quad T=\beta\,R^2 ,
\ee
where $R$ is the same as in (\ref{R-def}), and $\beta$ is an arbitrary number.  With these assumptions, one has the following simplifications
 \bea
\nnb   {\cal V} &=& (x_2)^3\, R^2 \quad ,\quad    {\cal V}_a \,=\,(x_2)^2 \, R\, e_a \quad,\quad     {\bf f }\,=\, x^3 R^2 \quad,\quad     {\bf f}_a \,=\, x^2  R\, e_a\,,\\ [2mm]
\nnb\tilde{\cal V} &=& (z_2)^3\, T \quad \;\;,\quad \tilde{\cal V}_i \,=\,(z_2)^2 \,  T_i \qquad\,,\quad  g \,=\, z^3\, T \quad\,,\quad g_i \,=\,z^2 \, T_i\, \\ [2mm]
W_{+} &=& 6\,R\,  x\, (z^2\beta R-1)\quad\;,\;\quad    \partial_a W_+ \,=\, 2(z^2 \beta R-1) e_{a} \quad\,,\quad  \partial_i W_+  \,=\,4\, x \, z\, T_{i} \nn\\ [2mm]
W_{3} &=& {\bf q}_0- 3\, R\, x\,  \hat\xi    + p^0 R^2 x^3  \quad\;,\;\quad \partial_a W_3\,=\, (\,p^0 R\,x^2 - \hat\xi\,)\,e_a  \,, 
\label{eq:QuantitiesManyHyp}  \eea
  where we introduced
  \be
  \hat\xi \,=\,\xi^0-\beta R \,\zeta \,.
  \ee
Now let us consider the first order conditions of subsection \ref{1stOrderCond}.
With respect to the case of a single hypermultiplet, here we have in addition the last equation in (\ref{eq:1stOrderAnsatz}). One can see that its solution requires $\beta>0$ for consistency with the assumption \hbox{$R>0$, and reads}
\be\label{eq:Solz}
z\,=\, -i\,\sqrt{\frac{3}{\beta R}}\;.
\ee
Plugging this into $W_{\pm}$ in (\ref{eq:QuantitiesManyHyp}), one gets $W_{\pm}=-24 x R$ and $D_a W_{\pm}=-8 e_a$. Comparing with (\ref{eq:Wansatz}), we see that after substituting (\ref{eq:Solz}) the Killing prepotentials ${\cal P}_x$ of this subsection are related to those of the previous subsection, here denoted by ${\cal P}_x|_{h_1=0}\,$, by
\be
{\cl P}_\pm  \,=\, \lambda \, {\cl P}_\pm|_{h_1=0}\qquad,\qquad {\cl P}_3  \,=\, {\cl P_3}|_{h_1=0\,,\,\xi\,\to\,\widehat\xi}
\ee          
where the proportionality factor
\be
\lambda \,=\, \frac{2}{3^{\frac{3}{4}}} \left( \frac{\beta}{R} \right)^{\frac{1}{4}}
\ee
arises from the different expressions for $e^{\frac{K_1}{2}}$ in the present ($h_1 > 0$) case with respect to the universal hypermultiplet (\hbox{$h_1= 0$}) case. It follows that the three solutions (\ref{sola})--(\ref{solc})  of the previous subsection generalize to the present case of an arbitrary number of vector and hypermultiplets. Labelling by $\ldots|_{h_1=0}$ the quantities appearing in (\ref{sola})--(\ref{solc}), we infer the solutions for the current $h_1>0$ case: 
\bea
x_1 \!\!&=&\!\! x_1|_{h_1=0}\qquad,\qquad x_2\,=\, x_2|_{h_1=0}\qquad,\qquad \hat \xi \,=\, \xi\,|_{h_1=0}\nnb \\ [2mm]
e^\varphi \!\!&=&\!\! \lambda\, e^{\varphi}|_{h_1=0} \,\sim\,   \left( {   R^{1\over 4} \beta^{3\over 4} \over   p^0 {\bf q}_0^2 }\right)^{1\over 3}\qquad\;\;,
\qquad\;\;     V  \,=\, \lambda^4\,V|_{h_1=0} \;\sim\;  \left( { R\, \beta^3\over p^0 {\bf q}_0^5 }\right)^{1\over 3},  \label{solgen}
\eea
with $z$ given by (\ref{eq:Solz}).

\subsection{\label{U-Invariant-Lambda}U-invariant cosmological constant}

In this section, we propose a U-duality invariant formula for the
dependence on NSNS and RR fluxes of the scalar potential $V$ at its
critical points. This defines the cosmological constant $\Lambda =\left. V\right| _{\partial V=0}\,$. 

As considered in the treatment above, the setup is the
following. The vectors' and hypers' scalar manifolds are given by
$\mathscr M_2 = G/H$ and $ {\mathscr M}_{Q}$. Here, $G/H$ is a \textit{symmetric} special K\"{a}hler
manifold with cubic prepotential ($d$-special K\"{a}hler space, see
\textit{e.g.} \cite {deWit:1992wf}), with complex dimension
$h_{2}$, coinciding with
the number of (abelian) vector multiplets. On the other hand, ${\mathscr M}%
_{Q} $ is a \textit{symmetric} quaternionic manifold, with
quaternionic dimension $h_{1}+1$, corresponding to the number
of hypermultiplets. The manifold ${\mathscr M}_{Q}$ is the 
c-map~\cite{Cecotti:1988qn} of the symmetric $d$-special K\"{a}hler space $%
\mathscr M_1 =\mathscr{G}/\mathscr{H}\subsetneq   {\mathscr M}_{Q}$, with complex dimension
$h_{1}$.
Thus, the overall U-duality group is given by\footnote{In this paper, we call U-duality group the symmetry group that has
a symplectic action on special K\"ahler manifolds $\mathscr M_2\times\mathscr M_1$, and not the whole symmetry group of the overall scalar manifold $\mathscr M_2\times\mathscr M_Q$.}
\begin{equation}
U\,\equiv\, G\times \mathscr{G}\,\subsetneq\, Sp\left(
2h_{2}+2,\mathbb{R}\right) \times Sp\left(
2h_{1}+2,\mathbb{R}\right) .
\end{equation}
The Gaillard-Zumino \cite{Gaillard:1981rj} embedding of $G$ and
$\mathscr{G}$ is provided by the symplectic representations
$
\mathbf{R}_{G}$ and $  
\mathbf{R}_{\mathscr{G}}\,$,  respectively spanned by the symplectic indices $\mathbb{A}$ and
$\mathbb{I}$.
The RR fluxes $c^{\mathbb{A}}= (p^0,p^a,q_0,q_a)$ sit in the $\left( 2h_{2}+2\right)$ vector representation $%
\mathbf{R}_{G}$, whereas the NSNS fluxes fit into the $\left( 2h_{2}+2\right) \times
\left(
2h_{1}+2\right) $ bi-vector representation $\mathbf{R}_{G}\times \mathbf{R}_{\mathscr{G}} $
\begin{equation}
Q^{\mathbb{AI}}=\mathbb{C}_{2}^{\mathbb{AB}}Q_{\mathbb{B}}^{~\mathbb{I}%
}
=
\left(
\begin{array}{cc}
  m^{AI} &   m^A{}_I    \\ [1mm]
   -e_A{}^{ I} &   -e_{A I}    \\
   \end{array}
\right).
 \label{Q-def}
\end{equation}
\textit{A priori}, in presence of $c^{\mathbb{A}}$ and
$Q^{\mathbb{AI}}$, various $\left( G\times \mathscr{G}\right)
$-invariants, of different orders in RR and NSNS fluxes, can be
constructed. Below we focus our analysis on invariants of
total order four and sixteen in fluxes, which respectively turn out
to be relevant for the U-invariant characterization of $\Lambda $
for the solutions \textbf{A}, \textbf{B}, \textbf{C} of subsections \ref{OneHyper} and \ref {Many-Hyperss}.

\subsubsection*{Only universal hypermultiplet}

Special K\"ahler symmetric spaces are characterized by a constant completely symmetric 
symplectic tensor $d_{\bb A_1 \bb A_2 \bb A_3 \bb A_4}$. This tensor  defines a quartic $G$-invariant
$\mathcal{I}_{4}\left( c^{4}\right)$  given by 
\begin{eqnarray}
\mathcal{I}_{4}\left( c^{4}\right)  &=&d_{\bb A_1 \bb A_2 \bb A_3 \bb A_4} c^{\bb A_1} \ldots c^{\bb A_4}\label{i4gen}
 \\ [2mm]
&=& -(p^0 q_0+p^a q_a)^2 +\ft{2}{3}d_{abc}\,q_{0}p^{a}p^{b}p^{c}- 
\ft{2}{3}d^{abc}\,p^{0} q_{a}q_{b}q_{c} +d_{abc}d^{aef}p^{b}p^{c}q_{e}q_{f}\,. 
 \label{I_4(c^4)} \nn \end{eqnarray}
The non-trivial components of $d_{\bb A_1 \bb A_2 \bb A_3 \bb A_4}$ are listed in (\ref{uinv-5}).
A similar definition holds for the symplectic tensor $d_{\bb I_1 \bb I_2 \bb I_3 \bb I_4}$ of the symmetric coset $\mathscr G /\mathscr H$.
 
For the explicit solutions found above, the RR fluxes $c^{\bb A}$ satisfy $p^{0}\neq 0$ and ${\bf q}_a=0$, i.e.
\begin{equation}
q_{a}\equiv \frac{1}{2}\frac{d_{abc}p^{b}p^{c}}{p^{0}}.  \label{13}
\end{equation}
Plugging this into (\ref{i4gen}) and  using the relation (\ref{idsym}) (holding in \textit{homogeneous symmetric} $d$-special K\"{a}hler
geometries) one finds
\be
\mathcal{I}_{4}\left( c^{4}\right)  \;=\; -\left( p^{0}\right)^{2}\mathbf{q}%
_{0}^{2} \qquad\qquad\qquad\quad \textrm{with}\qquad\mathbf{q}_{0} \,\equiv\, q_{0}+\frac{1}{6}\frac{d_{abc}p^{a}p^{b}p^{c}}{%
\left( p^{0}\right) ^{2}}\,.\label{q0-bold-def}
\ee
 Notice that the full dependence on $p^a$ is encoded in the shift $q_0 \to {\bf q}_0$ and therefore we can, 
without loosing in generality,  restrict ourselves to the simple choice  $c^{\bb A}=(p^0,0,{\bf q}_0,0)$.  
The RR and NSNS fluxes $c^{\bb A}$ and $Q^{\mathbb{AI}}$ are then given by
\begin{equation}
c^{\mathbb{A}} \equiv \left(
\begin{array}{c}
p^0 \\
0 \\
{\bf q}_0 \\
0
\end{array}
\right)\qquad, \qquad Q^{\mathbb{A}}{}_0\equiv \left(
\begin{array}{c}
0 \\
0 \\
0 \\
-e_{a}
\end{array}
\right)  \qquad, \qquad Q^{\mathbb{A}0}=0 \,.  \label{Q^A}
\end{equation}
 Besides ${\cal I}_4(c^4)$ one can build the following non-trivial quartic invariant
  \be \label{i4cq3}
  {\cal I}_4(c Q^3) =d_{\bb A_1 \bb A_2 \bb A_3 \bb A_4 } c^{\bb A_1} \,Q^{\bb A_2}{}_0\,Q^{\bb A_3}{}_0\,Q^{\bb A_4}{}_0=
  -\ft16 p^0 d^{abc} e_a e_b e_c= -p^0 R\,.
  \ee
Let us remark that ${\cal I}_4(c Q^3)$ is also invariant under the group $SO(2)=U(1)$
which, due to the absence of special K\"ahler scalars $z^i$ in the hypersector, gets promoted to global symmetry. $\mathscr G =SO(2)$ is
thus embedded into the symplectic group $Sp(2,\bb R)$ via its irrepr.
$\bf 2$, through which it acts on symplectic sections $(Z^0, \cl G_0)$. The
$SO(2)$-invariance of ${\cal I}_4(c Q^3)$ is manifest, because the latter depends on the $SO(2)$-invariant ${\cal I}_2\big((Q_{\bb A})^2\big)\,=\,(Q_{\bb A  0})^2  +(Q_{\bb A}{}^0)^2\,=\,(Q_{\bb A  0})^2 \;$ (no sum over $\bb A$ is understood).
  
It is easy to see that the expressions in eqs.\ (\ref{sola})--(\ref{solc})
 depend only on the two combinations ${\cal I}_4(c^4)$ and ${\cal I}_{4}(c Q^3)$ given in  (\ref{i4gen}) and  (\ref{i4cq3})
respectively. In particular, we obtain a manifestly
$(G\times U(1))$-invariant formula for the AdS cosmological constant  at the critical points
\begin{equation}
\Lambda=V  \,\sim\, -\frac{\mathcal{I}_{4}^{4/3}\left( cQ^{3}\right) }{%
\left| \mathcal{I}_{4}\left( c^{4}\right) \right| ^{5/6}} \,\sim\, {Q^4\over c^2} \label{n_H=1-Lambda}
\,.
\end{equation}
Notice that RR and NSNS\ fluxes play very different roles in their contribution to $%
\Lambda$. Indeed, the cosmological constant grows quartically on NSNS fluxes and fall off quadratically on RR charges. It would be nice
to understand whether this is a general scaling feature of the gauged supergravities under study.

\subsubsection*{Many hypermultiplets}

Next let us consider the case with arbitrary number of hypermultiplets. 
From (\ref{solgen}),  it follows that the cosmological constant $\Lambda=V$ in this case  depend only on the 
combinations  ${\cal I}_4=(p^0{\bf q}_0)^2$ and ${\cal I}_{16} \sim (p^0)^4 R \beta^3 \sim c^4 Q^{12} $.  
In order to write $\Lambda$ in a U-duality invariant form  we should then find an invariant ${\cal I}_{16}$ built out of 12 $Q$'s and 4 $c$'s that reduce to   $(p^0)^4 R \beta^3$ on our choice of RR and NSNS fluxes.  
 The following $\left( G\times \mathscr{G}\right)$-invariant quantity does the job
\begin{eqnarray}
\mathcal{I}_{16}\left( c^{4}Q^{12}\right) &\equiv &d_{\mathbb{I}_{1}\mathbb{I%
}_{2}\mathbb{I}_{3}\mathbb{I}_{4}}d_{\mathbb{I}_{5}\mathbb{I}_{6}\mathbb{I}%
_{7}\mathbb{I}_{8}}d_{\mathbb{I}_{9}\mathbb{I}_{10}\mathbb{I}_{11}\mathbb{I}%
_{12}}d_{\mathbb{A}_{1}\mathbb{B}_{1}\mathbb{B}_{2}\mathbb{B}_{3}}d_{\mathbb{%
A}_{2}\mathbb{B}_{5}\mathbb{B}_{6}\mathbb{B}_{7}}d_{\mathbb{A}_{3}\mathbb{B}%
_{9}\mathbb{B}_{10}\mathbb{B}_{11}}d_{\mathbb{A}_{4}\mathbb{B}_{4}\mathbb{B}%
_{8}\mathbb{B}_{12}} \times \notag \\ [2mm]
&&\times c^{\mathbb{A}_{1}}c^{\mathbb{A}_{2}}c^{\mathbb{A}_{3}}c^{\mathbb{A}_{4}}Q^{%
\mathbb{B}_{1}\mathbb{I}_{1}}\ldots 
Q^{\mathbb{B}_{12}\mathbb{I}_{12}} .
\label{I_16-our-case-def}
\end{eqnarray}
 The explicit expression of $\mathcal{I}%
_{16}\left( c^{4}Q^{12}\right) $ is rather intricate. Nevertheless, this formula undergoes
a dramatic simplification when considering the 
configuration of NSNS and RR fluxes supporting the solutions found in subsection \ref{Many-Hyperss}. As before we encode the full dependence on $p^a$  in the shift $q_0 \to {\bf q}_0$ and therefore we   
 restrict ourselves to the charge vector choice  $c^{\bb A}=(p^0,0,{\bf q}_0,0)$.  
More precisely, we take NSNS and RR fluxes with all components of $Q^{\mathbb{AI}}$, $c^{\bb A}$  zero 
except for 
\be
  Q_{a}^{~i} = -e_{a}{}^i\;, \qquad   Q_{a0}=-e_{a}\;,   \qquad c^0=p^0\;, \qquad c_0={\bf q}_0\,, \label{qrest}
\ee
where $e_{a}^{~i}$ satisfy the constraint (\ref{Constraintond}) for some $\beta \in
\mathbb{R}_{+}$.  A simple inspection to (\ref {I_16-our-case-def}),  shows that contributions to
  ${\cal I}_{16}$ come only from the components $d^0{}_{ijk}=-\ft16 \, d_{ijk}$ of 
$d_{\bb I_1 .. \bb I_4}$ and $d_0{}^{b_1 b_2 b_3}=-\ft16 d^{abc}$ of $d_{\bb A \bb B_1 \bb B_2 \bb B_3}\,$.  
 Indeed using
  (\ref{idsym})  one finds  (see appendix \ref{sAd} for details)
\begin{eqnarray}
\mathcal{I}_{16}\left( c^{4}Q^{12}\right) &=&  \gamma  \left( p^{0}\right)^{4} R\,\beta ^{3},  \label{9-simpl}
\end{eqnarray}
with\footnote{ Interestingly, the quantity $d^{abc}d_{abc}$, appearing in (\ref{s}) is related to the Ricci scalar
curvature $\mathcal{R}$ of the vector multiplets' scalar manifold
$G/H\,$, whose general expression for a $d$-special
K\"{a}hler space reads  $\mathcal{R}= -\left( h_{2}+1\right) h_{2}+d^{abc}d_{abc}$, see \cite{Cremmer:1984hc}.}
\bea
\gamma \equiv \frac{1}{ 6^{6}}  (d^{abc} d_{abc} + h_2+3)^3. \label{s}
\eea
We conclude that the cosmological constant of the AdS vacuum solutions 
obtained in subsection~\ref{Many-Hyperss} can be written in a manifestly U-duality invariant form in terms of ${\cal I}_4(c^4)$ and ${\cal I}_{16}(c^4 Q^{12})$, and reads
\begin{equation}
\Lambda\,=\,V \,\sim\, \frac{\mathcal{I}_{16}^{1/3}\left(
c^{4}Q^{12}\right) }{\left| \mathcal{I}_{4}\left( c^{4}\right)
\right| ^{5/6}}\,\sim\, {Q^4\over c^2} \,. \label{n_H>1-Lambda}
\end{equation}
\vskip 5mm $\:$


\subsection{Central charge via entropy function}\label{EntrFct}

 According to holography, gravity theories on AdS space are related to CFT's living on the AdS boundary. The central charge of the CFT can be extracted from the so called {\it entropy function} $F$
 evaluated at the near horizon geometry \cite{Ferrara:2008xz}.
For AdS$_2$, this function gives the entropy  of the black hole, for AdS$_3$ the Brown-Henneaux central charge. In general, $F$ computes the extreme value of the supergravity ``c-function" introduced in~\cite{Girardello:1998pd} (see also \cite{Freedman:1999gp}).  This quantity is a U-duality invariant and provides us with the basic macroscopic information about the boundary physics. In this subsection we apply the entropy function formalism to our AdS$_4$ solutions, and derive a U-duality invariant macroscopic formula for the central charge of the dual CFT$_3$.

The entropy function $F$ is defined as the Legendre transform with respect to the electric charges\footnote{We call electric the field-strengths filling the timelike direction.} of the higher-dimensional supergravity action evaluated at the solution. In our approach, we have dualized all electric charges (i.e. the 4--form flux along spacetime) into magnetic ones (i.e. internal fluxes), hence $F$ is simply minus the supergravity action.\footnote{ Alternatively, the same results can be found by considering instead of the 6--form along the internal space a 4--form electric flux along AdS$_4$, and performing the Legendre transform on this flux variable.} Reducing the higher-dimensional action down to four dimensions, taking the 4d scalars $\phi^i$ to be constant and all the other non-metric fields to vanish, one has
 \be
 F=-S_{\rm IIA} \,=\, -{1\over 2 \kappa^2}\int_{{\rm AdS}_4} \!\!\! d^4 x \sqrt{g_4 } \left[ R_4- 2V(\phi^i) \right] \,=\, - r_{\rm AdS}^4  \left[ R_4 -2V(\phi^i) \right],
 \label{ffunc}
 \ee
where $R_4$ is related to the AdS$_4$ radius $r_{\rm AdS}$ by $R_4=-{12\over r_{\rm AdS}^2 }$, and we regularize the infinite volume of AdS$_4$ in such a way that
  \be
 r_{\rm AdS}^4 = {1\over 2 \kappa^2} \int_{{\rm AdS}_4} \!\!\!d^4 x \sqrt{g_4 }\,.
 \ee
  Any other choice of normalization redefines $F$ by an irrelevant flux-independent constant.
 The supergravity vacuum follows then by extremizing $F$ with respect to the scalar fields $\phi^i$ and the AdS radius $r_{\rm AdS}\,$:
 \be
 {\partial F\over \partial \phi^i}={\partial F\over \partial r_{\rm AdS}}=0\,.
 \ee
 Denoting by $\langle\phi\rangle$ a solution of  $\pd{V}{\phi^i}=0$, one finds for the
 AdS radius $r_{\rm AdS}$ and the central charge $F$
 \bea
    F=6 \,r_{\rm AdS}^2  =-\frac{18}{V(\langle\phi\rangle)} \label{jazz}\,.
 \eea
  The right hand equation reproduces the Einstein equation that relates the AdS radius $r_{\rm AdS}$ (or, equivalently, the cosmological constant $\Lambda = -3/ r_{\rm AdS}^2$) to the vev of the scalar potential. Notice that solutions make sense only for $\langle V\rangle <0$.

Taking into account the results of subsections \ref{OneHyper} and \ref{Many-Hyperss},
as well as eqs.\ (\ref{n_H>1-Lambda}) and (\ref{n_H=1-Lambda}), one
obtains the U-invariant expressions of the central charge $F$,
respectively for the case with only
universal hypermultiplet and many hypermultiplets  
\begin{equation}
F_{h_{1}=0}\,\sim\, \frac{\left| \mathcal{I}_{4}\left( c^{4}\right)
\right| ^{5/6}}{\mathcal{I}_{4}^{4/3}\left( cQ^{3}\right)
}\quad\Rightarrow\quad  F_{h_{1}=0}\,\sim\,\frac{ c^{2}}{ Q^{4}}\,. \label{F-univ-hyper}
\end{equation}
\begin{equation}
F_{h_{1}>0}\,\sim\, -\frac{\left| \mathcal{I}_{4}\left( c^{4}\right)
\right| ^{5/6}}{\mathcal{I}_{16}^{1/3}\left( c^{4}Q^{12}\right)
}\quad\Rightarrow\quad   F_{h_{1}>0} \,\sim\, \frac{ c^{2}}{Q^{4}}\;, \label{F-many-hypers}
\end{equation}
Notice the different contribution of RR and NSNS fluxes to the
central charge $F$. Analogously to the black hole entropy, the
central charge grows quadratically in the RR charges (RR fluxes),
but falls off quartic in the NSNS charges ($H$-flux and (non-)geometric fluxes). The same scaling behaviour was found in \cite{Aharony:2008wz} for the central charge of the CFT$_3$ dual to type IIA on the AdS$_4\times T^6/\bb Z^2_3$ orientifold background. It would be interesting to understand to what extent this is a general feature of flux compactifications.

      
\subsection{Examples from type IIA/IIB compactifications}\label{StringRealizations}

Explicit examples of compact manifolds $M_6$ yielding $\cl N=2$ supergravity upon dimensional reduction of type II theories are the cosets displayed in table \ref{scalarmanifolds}. Type IIA reductions on these spaces have been studied in \cite{Cassani:2009ck}; see also \cite{House:2005yc,KashaniPoorNearlyKahler}, and \cite{Caviezel:2008ik} for the relative $\cl N=1$ orientifold truncations. The cosets $M_6$ of table \ref{scalarmanifolds}  admit an $SU(3)$ structure (see appendix~\ref{ScalPotIIA} for a definition), and correspond respectively to the sphere $S^6$, the complex projective space $\bb{CP}^3$, and the flag manifold $\bb F(1,2;3)$, endowed with a left-invariant metric.\footnote{The flag manifold $\bb F(1,2;3)$ is defined as the set of pairs made by a line and a plane in $\bb C^3$ such that the line belongs to the plane.} On these spaces, the left-invariant metric and B-field deformations span a special K\"ahler manifold. In type IIA reductions, the latter corresponds to the vector multiplet scalar manifold, and for the three cosets at hand one obtains respectively the $t^3$, the $st^2$ and the $stu$ models of $\cl N=2$ supergravity. In addition, the compactification yields the universal hypermultiplet $(a,\varphi,\xi^0,\tilde\xi_0)$, parameterizing the quaternionic manifold $\frac{SU(2,1)}{U(2)}$, whose isometries are gauged as described in section \ref{RevisitingPotential} and appendix \ref{saltd}.  
  
\begin{table}
\begin{center}
\begin{tabular}{|c|c|c|c|}
\hline
\rule{0pt}{4ex}
        $M_6\qquad$ &  $\frac{G_2}{SU(3)}= S^6$ &  $\frac{Sp(2)}{S(U(2)\times U(1))}= \bb{CP}^3$  & $\frac{SU(3)}{U(1)\times U(1)}= \bb F(1,2;3)$ \\[3mm]
\hline
\rule{0pt}{9ex}
Type IIA $\left\{\begin{array}{c} {\mathscr M}_2 \,\leadsto f \\ [4mm]  {\mathscr M}_1 \\ [4mm] {\mathscr M}_Q \end{array}\right.$ & $\begin{array}{c}
\frac{ SU(1,1)}{ U(1)}\,\leadsto t^3\\ [5mm]   -- \\ [5mm] \frac{SU(2,1)}{U(2)} \end{array}$ & $\begin{array}{c} \Big(\frac{SU(1,1)}{U(1)}\Big)^2\,\leadsto st^2 \\ [4mm] -- \\ [5mm] \frac{SU(2,1)}{U(2)} \end{array}$ &
 $\begin{array}{c} \Big(\frac{SU(1,1)}{U(1)}\Big)^3\,\leadsto stu \\ [5mm] -- \\ [5mm] \frac{ SU(2,1)}{U(2)} \end{array}$\\ [2mm]
\hline
\rule{0pt}{10ex}
Type IIB $\left\{\begin{array}{c} 
{\mathscr M}_2  \\ [4mm] {\mathscr M}_1 \,\leadsto g  \\ [4mm] {\mathscr M}_Q \end{array}\right.$ & $\begin{array}{c} -- \\ [4mm]\frac{SU(1,1)}{U(1)}\,\leadsto t^3\\ [5mm]  \frac{G_{2(2)}}{ SO(4)} \end{array}$ & $\begin{array}{c} -- \\ [3mm]\Big(\frac{ SU(1,1)}{ U(1)}\Big)^2 \,\leadsto\,st^2\\ [5mm] \frac{ SO(4,3)}{ SO(4)\times SO(3)} \end{array}$ &
 $\begin{array}{c}
 --\\ [4mm]\Big(\frac{ SU(1,1)}{ U(1)}\Big)^3\,\leadsto\,stu \\ [5mm] \frac{SO(4,4)}{ SO(4)\times SO(4)}  \end{array}$
\\ [2mm]\hline
\end{tabular}
\caption{Supergravity scalar manifolds arising from $\cl N=2$ compactifications of type II theories on the 6d coset manifolds $M_6$. We recall that $\mathscr M_2$ is the special K\"ahler manifold parameterized by the scalars in the vector multiplets, while $\mathscr M_1$ is the special K\"ahler submanifold of the quaternionic hyperscalar manifold $\mathscr M_Q$, determining the latter via \hbox{c-map}.
We also display the form of the prepotentials $f$ and $g$ associated respectively with $\mathscr M_2$ and $\mathscr M_1$.} \label{scalarmanifolds}
\end{center}
\end{table}

One can also consider type IIB compactified on the same cosets, again implementing a left-invariant reduction ansatz. In this case, one gets a 4d $\cl N=2$ supergravity with no vector multiplets and a number of hypermultiplets going from 2 to 4, depending on the chosen $M_6$. The special K\"ahler coset moduli space, which for type IIA compactifications was identified with the vector multiplet scalar manifold, in type IIB corresponds to a submanifold of the hyperscalar quaternionic manifold, and determines the latter via the c-map. The additional coordinates are given by the axion dual to the $B$-field in 4d, by the 4d dilaton $\varphi$, and by the scalars coming from the expansion of the type IIB Ramond-Ramond potentials $C_0$, $C_2$ and $C_4$ in a basis of left-invariant forms of even degree on $M_6$. For instance, for a compactification on $\frac{G_2}{SU(3)}= S^6$, the quaternionic manifold is the eight-dimensional non-compact coset $\frac{G_{2(2)}}{ SO(4)}$, corresponding to the image of $\frac{SU(1,1)}{U(1)}$ under the c-map \cite{BodnerCadavid}. Table \ref{scalarmanifolds} collects the various scalar manifolds arising in these type IIA/IIB coset dimensional reductions.

For type IIA coset compactifications, the $\cl N=2$ scalar potential and its vacuum structure were studied in \cite{Cassani:2009ck}. The further results obtained above --- specifically, the first order equations of subsection \ref{1stOrderCond}, and the U-invariant formulae of subsections \ref{U-Invariant-Lambda}, \ref{EntrFct} for the case of the universal hypermultiplet --- hold in particular for these coset examples. Moreover, for the cosets $\frac{Sp(2)}{S(U(2)\times U(1))}$ and $\frac{SU(3)}{U(1)\times U(1)}$, the explicit solutions of subsection~\ref{OneHyper} extend those given in \cite[sect.$\:$6]{Cassani:2009ck} in that they allow for non-vanishing $p^a$ and $q_a$, corresponding to fluxes of the $G_2$ and $G_4$ RR field-strengths (cf.~appendix \ref{ScalPotIIA}). Thanks to the consistency of the reduction, which also was established in \cite{Cassani:2009ck}, the AdS vacua lift to {\it bona fide} solutions of type IIA supergravity. 

Concerning type IIB, the reduction based on a left-invariant ansatz can still be shown to be consistent using the same arguments as for type IIA, and the scalar potential is a subcase of the general formula (\ref{pot0}) as well. However, this IIB scalar potential turns out to be less interesting than for type IIA, since it displays a runaway behaviour. Indeed, this case falls in the situation, considered in~\cite{D'Auria:2007ay}, in which one has more hyper than vector multiplets, and the rank of the matrix $Q$ is $h_2+1$ ($=1$ here): then the equations of motion for the RR scalars $\xi^{\bb I}$ imply $(c+ \widetilde Q\xi)= 0$, which in turn sets $V_{\rm R} =0$. This leaves us with an effective scalar potential $V=V_{\rm NS}$, which is runaway due to the overall $e^{2\varphi}$ factor.

An issue that is left open is whether it is possible to identify the type IIB mirrors of type IIA compactified on the cosets $M_6$ above.

\newpage


\section{Discussion}\label{disc}

In this paper we studied the scalar potential of ${\cal N}=2$ gauged supergravities underlying flux compactifications of type IIA/IIB theories.
Exploiting the ${\cal N}=2$ formalism --- in particular, the special K\"ahler geometry on the scalar manifolds $\mathscr M_1$ and $\mathscr M_2$ --- we have written the scalar potential and its extremization conditions in terms of the Killing prepotentials $\cl P_x$ and their special K\"ahler
covariant derivatives. The equations for AdS vacua are solved via the system of first order conditions given in section \ref{1stOrderCond}, accounting for both supersymmetric and non-supersymmetric solutions. This first order ansatz may be thought as a possible alternative to the method of the ``fake superpotential"~\cite{Ceresole:2007wx} in the search
for a unifying principle to describe extremal solutions in supergravity. It would be interesting to study the lifting of our ansatz to a ten-dimensional context, where it corresponds to a deformation (by means of the parameters $u,v,\theta$) of the pure spinor equations for $\cl N=1$ backgrounds derived in \cite{GMPT12,GMPT3} employing the methods of generalized complex geometry. In this perspective, it would also be interesting to investigate the possible relations with the approach to ``partially BPS vacua'' developed in \cite{Lust:2008zd}. Notice that also our first order ansatz leaves unbroken a subset of the supersymmetry conditions (namely, the last line of (\ref{eq:1stOrderAnsatz})).

We found three (one supersymmetric and two non-supersymmetric) infinite series of AdS$_4$ vacua with a rich pattern of NSNS and RR fluxes. These generalize the solutions of \cite{Cassani:2009ck}, derived from flux compactifications on cosets with $SU(3)$ structure, to the case
of a general ${\cal N}=2$ cubic, symmetric supergravity with an arbitrary number of vector and hypermultiplets. We remark that, having these solutions as a starting point, one can generate full orbits of solutions by acting on the charges and on the symplectic sections of $\mathscr M_1$ and $\mathscr M_2$ with U-duality transformations. We leave for the future a detailed analysis of these orbits, as well as the search for the precise 10d lifting of the full set of our solutions. A further issue to be investigated concerns the stability of the $\mathcal N=0$ AdS vacua: to conclude about this, one should test the Breitenlohner-Freedman bound on the solutions, as done in \cite{Cassani:2009ck} within a theory including the universal hypermultiplet and at most 3 vector multiplets. It might be that our first order relations turn out to be useful at this scope. It would also be interesting to go beyond our classical analysis and study quantum effects in this context, possibly combining 4-dimensional and string compactification methods.

Finally, we proposed a U-duality invariant formula for the cosmological constant built out of the  NSNS charges $Q_{\!\bb A}^{\;\;\bb I}$, the RR charges $c^{\bb A}$, and the characteristic quartic tensors $d_{{\bb I_1}\ldots{\bb I_4}}$,   $d_{{\bb A_1}\ldots{\bb A_4}}$. This invariant also describes the central charge of the dual 3-dimensional CFT, as follows from explicit evaluation of the entropy function on the AdS$_4$ solution.

\vskip 8mm

\noindent
{\large \bf Acknowledgments}\vskip 2mm

\noindent We would like to thank Lilia Anguelova and Gianguido Dall'Agata for useful discussions, as well as Paul Koerber for interesting correspondence. This work is supported in part by the ERC Advanced Grant no.~226455, \textit{%
``Supersymmetry, Quantum Gravity and Gauge Fields''}
(\textit{SUPERFIELDS}). D.C. thanks the Rome ``Tor Vergata'' string theory group for support during his visit under the grant PRIN 2007-0240045. A.M. would like to thank the Department of Physics and Astronomy, UCLA, CA USA, where part of this work was done, for kind hospitality
and stimulating environment. The work of D.C. has been supported by the Fondazione Cariparo
Excellence Grant {\em String-derived supergravities with branes and
fluxes and their phenomenological implications}. The work of S.F.~has been supported in part by D.O.E.~grant DE-FG03-91ER40662,
Task C. The work of A.M. has been supported by an INFN visiting
Theoretical Fellowship at SITP, Stanford University, Stanford, CA
USA. The work of H.S. has been supported in part by the Agence
Nationale de la Recherche (ANR).


\begin{appendix}

\section{An alternative derivation of $V$}
\label{saltd}

In the following, first we discuss the general form of the scalar potential in $\cl N=2$ supergravity, then we provide an alternative way to derive expression (\ref{potid}) for the flux-generated potential.

In any theory of extended supergravity, a general Ward identity implies that the scalar potential $V$ is determined by the squares of the shifts that the gaugings induce in the fermionic susy transformations. In the $\cl N=2$ context, this Ward identity reads
\be\label{eq:WardIdentity}
V \delta^{\cl A}_{\;\;\cl B}\;=\; \;-\;12 \overline S^{\cl C\cl A} S_{\cl C\cl B} \;+\; g_{a\bar b}\,W^{a{\cl C\cl A}}\, W^{\bar b}_{\cl C\cl B}\; +\; 2N_{\bb I}^{\cl A}\, N^{\bb I}_{\cl B} \,,
\ee
where $\cl A, \cl B, \cl C =1,2$ are $SU(2)$ R-symmetry indices, and the matrices $S_{\cl A \cl B}$, $W^{a{\cl A\cl B}}$ and $N_{\bb I}^{\cl A}$ are the fermionic shifts appearing respectively in the supersymmetry transformations of the gravitini $\psi_{\mathcal A \mu}$, gaugini $\lambda^{a\mathcal A}$ and hyperini $\zeta_{\bb I}$:
\begin{eqnarray}
\nnb  \delta\psi_{\mathcal A\mu}  &= & \ldots + \nabla_\mu\ep_{\mathcal A} - S_{\mathcal A\mathcal B}\gamma_\mu\ep^{\mathcal B} \\ [1mm]
\nnb \delta\lambda^{a\mathcal A} & = & \ldots + W^{a\mathcal A\mathcal B}\ep_{\mathcal B}\\ [1mm]
\label{eq:N=2fermionicVar} \delta\zeta_{{\bb I}} & = &  \ldots + N_{{\bb I}}^{\mathcal A}\ep_{\mathcal A}\,.
\end{eqnarray}
A prominent role is played by the gravitino shift $S_{\cl{AB}}$, which is expressed in terms of the triplet of Killing prepotentials $\cl P_x = e^{\frac{K_2}{2}}(\cl P_{xA} X^A - \tilde{\cl P}{}_x^{A}\cl F_A)$, with $x=1,2,3$, encoding the gauging of the isometries in the hyperscalar manifold. Introducing as in the main text $\cl P_\pm = \cl P_1 \pm i\cl P_2$, one has the relation
\be\label{eq:GenFormS}
S_{\cl A\cl B}  =  \frac{i}{2}\sigma_{x\cl A\cl B}\, \cl P_x= -\frac{i}{2}
\left(
\begin{array}{cc}
  -\cl P_- &   \cl P_3   \\
   \cl P_3   &   \cl P_+ \\
\end{array}
\right),
\ee
where $(\sigma_x)_{\!\cl A}^{\;\;\cl B}$ are the standard Pauli matrices, and the $SU(2)$ index $\cl A=1,2$ is lowered with the antisymmetric tensor $\epsilon_{\cl A\cl B}$, using a SW-NE convention, i.e. $\sigma_{x\cl A\cl B} = \epsilon_{\cl B\cl C}(\sigma_x)_{\!\cl A}^{\;\;\cl C}$. We also raise the index with $\epsilon^{\cl A\cl B}=-\epsilon^{\cl B\cl A}$, satisfying $\epsilon^{\cl A\cl C}\epsilon_{\cl C\cl B}= -\delta^{\cl A}_{\,\cl B}$, hence $\sigma_x^{\;\cl A\cl B} = (\sigma_x)_{\!\cl C}^{\;\;\cl B}\epsilon^{\cl C\cl A}$.

Given the gravitino shift $S_{\cl{AB}}$, one has that the gaugino shift $W^{a\mathcal A\mathcal B}$ and the hyperino shift $N_{\bb I}^{\mathcal A}$ are determined by the derivatives of the $\cl P_x$ via \cite{Andrianopoli:1996cm, D'Auria:2001kv}
\begin{eqnarray}\label{eq:GaugMassMatrix} W^{a{\mathcal A\mathcal B}}& = &  i\sigma_x^{\,\cl A\cl B}  g^{a\bar b} D_{\bar b} \overline{\mathcal P}{}_x\\ [2mm]
\label{eq:HypMassMatrix} N_{\bb I}^{\mathcal A} & = & -\frac{1}{3}\,\cl U_{\;\;{\bb I} u}^{\mathcal A} \,\Om_x^{\,uv}\mathscr D_v\overline{\mathcal P}{}_x \;=\; 2\cl U_{\;\;{\bb I} u}^{\mathcal A}\bar k^u\,,
\end{eqnarray}
where the index $u$ labels the coordinates of the quaternionic manifold, and $\cl U_{\;\;{\bb I} u}^{\mathcal A}$ are the quaternionic vielbeine prior the dualization of a subset of the hyperscalars to tensor fields (cf.~footnote~\ref{PrecisationMagneticGaugings}). Furthermore, we put
\be\label{eq:Defk^u}
k^u= e^{\frac{K_2}{2}}(k_A^u X^A - \tilde k^{uA} \cl F_A),
\ee
where $k^u_A$ are the Killing vectors generating the quaternionic isometries being gauged with respect to electric gauge potentials, and and $\tilde k^{uA}$ are their magnetic counterparts. Finally, the covariant derivatives are defined by
\bea
D_a\cl P_x &=& (\partial_a + \ft12\partial_a K_2) \cl P_x\\ [2mm]
\label{eq:D_uP} \mathscr D_u \cl P_x &=& \partial_u \cl P_x + \epsilon_{xyz} (\om_y)_u \,\cl P_z = 2(\Om_x)_{uv} \,k^v\,.
\eea
Here $(\Om_x)_{uv}$ is the curvature 2--form of the hyperscalar quaternionic manifold, $\Om_x = (\Om_x)_{uv}dq^u\wedge dq^v$. Given the isometries $k^u$, relation (\ref{eq:D_uP}) actually {\it defines} the $\cl P_x$ in ${\cal N}=2$ supergravity.

Substituting (\ref{eq:GenFormS})--(\ref{eq:HypMassMatrix}) in (\ref{eq:WardIdentity}) and tracing over the $SU(2)$ indices $\cl A, \cl B$, one obtains the following standard expression for the $\cl N=2$ scalar potential \cite{Andrianopoli:1996cm, N=2withTensor1, N=2withTensor2}:
\be\label{eq:StandFormV}
V \; = \;  4 h_{uv}k^u \bar{k}^v  + \sum_x \left( g^{a\bar b}D_a\cl P_x D_{\bar b}\overline{\cl P}{}_x  -3 |\cl P_x|^2\right)\,.
\ee
This was the starting point adopted in \cite{D'Auria:2007ay} to obtain expression (\ref{pot0}).
Recalling (\ref{eq:D_uP}) and using the identity~\cite{Andrianopoli:1996cm}
\be
h^{uv}(\Om_x)_{su}(\Om_y)_{vt} = -\delta_{xy} h_{st} -\epsilon_{xyz}(\Om_z)_{st}\,,
\ee
we observe that (\ref{eq:StandFormV}) can also be recast in the following form, involving just the $\cl P_x$ and their derivatives\footnote{Notice that $h^{uv}\mathscr D_{u}\cl P_1 \mathscr D_v \overline{\cl P}{}_1 = h^{uv}\mathscr D_{u}\cl P_2 \mathscr D_v \overline{\cl P}{}_2 = h^{uv}\mathscr D_{u}\cl P_3 \mathscr D_v \overline{\cl P}{}_3
 $. }
\be
\label{eq:NewFormV2} V\;=\; \sum_{x=1,2,3} \left( \textstyle{\frac{1}{3}} h^{uv}\mathscr D_{u}\cl P_x \mathscr D_v \overline{\cl P}{}_x + g^{a\bar b}D_a\cl P_x D_{\bar b}\overline{\cl P}{}_x  -3 |\cl P_x|^2\right).
\ee
For our purposes, however, it is more convenient not to trace over $\cl A, \cl B\,$: we will instead consider the equivalent expression defined by taking $\cl A = \cl B = 2$ in (\ref{eq:WardIdentity}). This reads\footnote{One can also see that all the remaining information contained in (\ref{eq:WardIdentity}) amounts to the constraint $2(\Om_x)_{uv}\bar k^u k^v =  \epsilon_{xyz}\overline{\cl P}{}_y \cl P_z ,$ which is the abelian version of eq.~(7.56) in \cite{Andrianopoli:1996cm}, and has to be automatically satisfied for consistency.}
\be\label{eq:WardWithA=B=2}
V = g^{a\bar b}D_{a}\cl P_+ D_{\bar b} \overline{\cl P}{}_+ + g^{a\bar b}D_{a}\cl P_3 D_{\bar b} \overline{\cl P}{}_3
\, + \,\textstyle{\frac{1}{2}}\, h^{uv}\mathscr D_{u}\cl P_+ \mathscr D_v \overline{\cl P}{}_+  - \,3|\cl P_+|^2 - \,3 |\cl P_3|^2.\;\;
\ee
The two above expressions for the $\cl N=2$ scalar potential hold for any gauging involving just quaternionic isometries. We now evaluate the term in (\ref{eq:WardWithA=B=2}) containing the quaternionic covariant derivatives by specializing to the case of {\it dual} (also named {\it special}) quaternionic manifolds, which arise in Calabi-Yau \cite{Cecotti:1988qn, Ferrara:1989ik} and generalized geometry compactifications of type II theories. In this case, the quaternionic metric $h_{uv}$ is 
\be\label{eq:DualQuatMetric}
 h_{uv}d q^u d q^v \;=\;    g_{i\bar\jmath}\,d z^id \bar z^{\bar\jmath} + (d\varphi)^2 + \frac{e^{4\varphi}}{4}\big(d a - \xi^T\C_1 d \xi \big)^2 - \frac{e^{2\varphi}}{2} d \xi^T \bb M_1d \xi\,,
\ee
where $g_{i\bar\jmath}= \partial_{z^i}\partial_{\bar z^{\bar \jmath}}K_1$ is the metric on the special K\"ahler submanifold $\mathscr M_1$ of the dual quaternionic manifold. We also need the $Sp(1)$ connection 1--forms $\om_x$, which read \cite{Ferrara:1989ik}
\begin{eqnarray}
\nnb\om_1+i \om_2 &=& 2e^{\varphi} \, \Pi_1^T \C_1d \xi  \\ [2mm]
\label{SU2Connection}\om_3 &=& -\frac{e^{2\varphi}}{2} \left(d a -    \xi^T\C_1 d \xi \right)
+ \mathcal Q \,,
\end{eqnarray}
where $\cl Q$ is the $U(1)$ connection associated with the special K\"ahler geometry on $\mathscr M_1$ \cite{Andrianopoli:1996cm}:
\be
\mathcal Q\, =\,\ -\frac{i}{2}\big (\partial_i K_1 d z^i - \partial_{\bar \imath} K_1d \bar z^{\bar \imath}\big )
\,=\, -\frac{i}{2}\frac{Z^I\im\mathcal G_{IJ} d\bar Z^J - c.c.}{\bar Z^K\im\mathcal G_{KL} Z^L}\,.
\ee
The abelian isometries of the quaternionic metric (\ref{eq:DualQuatMetric}) that are gauged are generated by the Killing vectors
\be\nnb
\!\!\!\!\!\!\!\!\!\!\! k_A = -\big[2q_A +  e_A^{\;\bb I}(\C_1\xi)_{\bb I}\big]\pd{\,}{a}-   e_A^{\;\bb I}\,\pd{\,}{\xi^{\bb I}} \qquad,\qquad\tilde k^A = \big[\!-2p^A + m^{A\bb I}(\C_1\xi)_{\bb I}\big]\pd{\,}{a} + m^{A\bb I}\pd{\,}{\xi^{\bb I}}\,,
\ee
where $e_{\!A}{}^{\bb I} = (e_{\!A}{}^{I},e_{AI})^T\,$, $m^{A\bb I} = (m^{AI},m^A{}_{I})^T$, and the abelianity $[k_A,k_B]=[\tilde k^A, \tilde k^B]=[k_A, \tilde k^B]=0$ is ensured by (\ref{const}). Recalling (\ref{DefCharges}), (\ref{DefTildeQ}), the quantity in (\ref{eq:Defk^u}) then reads
\be\label{eq:k^u}
k = -\Pi_2^T \bb C_2(2 c + \widetilde Q\xi ) \pd{\,}{a} - \Pi_2^T Q \pd{\,}{\xi}\,.
\ee
The Killing prepotentials $\cl P_x$ associated with these isometries are given by \cite{Michelson, D'Auria:2007ay}
\be
\cl P_x = (\om_x)_u k^u.
\ee
By plugging in (\ref{SU2Connection}), (\ref{eq:k^u}), one finds the expressions given in (\ref{eq:KillingPrep}).
In the context of flux compactifications, the $\cl P_x$ can be derived by reducing the higher-dimensional gravitino transformation \cite{GLW1, GLW2}.

Then the quaternionic covariant derivatives $\mathscr D_u\cl P_x \equiv \partial_u \cl P_x + \epsilon_{xyz}(\om_y)_u\cl P_z$ read
\bea
\nnb \mathscr D_{z^i} \cl P_+ = (\partial_{i} + {\textstyle \frac{1}{2}}\partial_{i}K_1 )\cl P_+\quad&,&\quad\mathscr D_\varphi \cl P_+ = \cl P_+\\ [2mm]
\nnb\mathscr D_{(a)} \cl P_+ =  - \ft i2 e^{2\varphi}\cl P_+\;\;\qquad\qquad &,&\quad
\mathscr D_{\xi^\bb I} \cl P_+ = 2i e^{\varphi}\cl P_3 (\C_1\Pi_{1})_{\bb I} - \ft i2 e^{2\varphi} \cl P_+ (\C_1\xi)_{\bb I}\,.
\eea
The writing $\mathscr D_{(a)}$ emphasizes that here $a$ denotes the axion dual to the B--field, and should not be confused with a special K\"ahler index.

Also evaluating the inverse of (\ref{eq:DualQuatMetric}), we finally arrive at
\be\nnb
\!\!\!{\textstyle \frac{1}{2}} h^{uv}\mathscr D_{u}\cl P_+ \mathscr D_v \overline{\cl P}{}_+\,=\,
g^{i\bar\jmath} D_i\cl P_+ D_{\bar\jmath}\overline{\cl P}{}_+  + |\cl P_+|^2 +4 |\cl P_3|^2.
\ee
Substituting this into (\ref{eq:WardWithA=B=2}) we obtain precisely expression (\ref{potid}) for the scalar potential.

\section{Elaborating the $\cl N=1$ susy conditions}\label{SummaryN=1}

In this appendix we work out the $\cl N=1$ supersymmetry conditions within the $\cl N=2$ theory under study, building on the analysis done in \cite{Cassani:2007pq}.

\subsection{In terms of (derivatives of) the prepotentials $\cl P_x$}\label{N=1asDP}

We impose the vanishing of the $\cl N=2$ fermionic shifts given in (\ref{eq:N=2fermionicVar}) under a single supersymmetry parameter $\varepsilon$. The latter is related to the (positive chirality) $\cl N=2$ supersymmetry parameters  $\varepsilon_1$ and $\varepsilon_2$ via $\varepsilon_1 = a\varepsilon $ and $\varepsilon_2 = b\varepsilon $, with
\be\nnb
a = |a|e^{i\alpha}\qquad,\qquad b= |b|e^{i\beta}\qquad,\qquad |a|^2+|b|^2 =1\,.
\ee
The supersymmetry parameter $\ep$ is chosen to satisfy the Killing spinor equation on AdS$_4$: $\nabla_\nu \ep=
\frac{1}{2}\mu \gamma_\nu \ep^*$,
whose complex parameter $\mu$ is hence related to the AdS cosmological constant $\Lambda$ by $\Lambda = -3 |\mu|^2$.

The general $\cl N=1$ supersymmetry conditions can be condensed in the following linear equations for the $\cl N=2$ Killing prepotentials $\cl P_x$ introduced in (\ref{eq:KillingPrep}) and their covariant derivatives:
\bea
\label{eq:N=1eqsAppendix1} \mu(\,|a|^2 -|b|^2\,) &=& 0\quad \Rightarrow \quad |a|\,=\,|b| \;\; \textrm{in AdS}\\ [2mm]
\label{eq:N=1eqsAppendix2} \pm\, e^{\pm i\gamma}\cl P_\pm &=& 2\cl P_3 \;=\; -2i  \hat{\bar \mu}\, \\ [2mm]
\label{eq:N=1eqsAppendix3} \pm\, e^{\pm i\gamma}D_a\cl P_\pm &=& D_a\cl P_3\\ [2mm]
\label{eq:N=1eqsAppendix4} D_i \cl P_+ &=&  0\;\; =\;\; D_{\bar \imath} \cl P_-\,,
\eea
where we introduced $\gamma =\alpha-\beta+\pi$ and $\hat \mu = e^{-i(\alpha+\beta)}\mu$. The AdS condition $|a|=|b|$ is understood in (\ref{eq:N=1eqsAppendix2})--(\ref{eq:N=1eqsAppendix4}). The derivation of the above equations is a variation of the analysis done in \cite[sect.\:4]{Cassani:2007pq}. Upgrading the notation to the current conventions, in the following we write the equations given therein, corresponding to the vanishing of the fermion variations. The gravitino equation $\langle \delta_\ep\psi_{\mu\cl A}\rangle = 0$ yields
\bea\label{eq:CondOnSExplicit}
\nnb -\bar a\cl P_- + \bar  b\cl P_3 & = & ia\bar \mu \\ [2mm]
\bar a\cl P_3 + \bar b\cl P_+  &=& ib\bar \mu\,,
\eea
while the hyperino equation $\langle\delta_\ep\zeta_{\bb I}\rangle = 0$ gives
\begin{eqnarray}
\label{eq:HypVar1} 2 \bar a \mathcal P_3 + \bar b\cl P_+ &=& 0\\ [2mm]
\label{eq:HypVar2} \bar a \cl P_- -2\bar b \mathcal P_3 &=& 0\\ [2mm]
\label{eq:HypVar3} \bar b\, P_I^{\,\ul l}\, (\im \mathcal G)^{-1\,IJ}\big(Q_{J\bb A} - \mathcal N_{1\,JK} Q^K_{\;\;\,\bb A}  \big) \Pi_2^{\bb A} &=& 0\\ [2mm]
\label{eq:HypVar4} \bar a\,\overline P_I^{\,\ul{\bar l}}\,(\im \mathcal G)^{-1\,IJ}\big(Q_{J\bb A} - \overline{\mathcal N}_{1\,JK} Q^K_{\;\;\,\bb A}  \big) \Pi_2^{\bb A} &=& 0\,,
\end{eqnarray}
with $P_I^{\;\ul j}=(\,P_0^{\;\underline{j}},P_i^{\;\underline{j}}\,)=(-e_i^{\; \underline{j}}Z^i, e_i^{\; \underline{j}}\,)$, where $e_i^{\; \underline{j}},\;(i,j=1,\ldots,h_1)$ are the vielbeine of the special K\"ahler manifold $\mathscr M_1$ (the flat indices are underlined, and the choice of special coordinates $Z^I=(1,z^i)$ is understood). Finally, the gaugino equation $\langle \delta_\ep \lambda^{a\cl B}\rangle = 0$ is
\bea\label{eq:GauginoEqOriginal}
\nnb\bar a D_a \cl P_- - \bar b D_a \cl P_3 &=& 0\\ [2mm]
\bar a D_a \cl P_3 + \bar b D_a \cl P_+ &=& 0\,.
\eea
Now it is easy to see that (\ref{eq:CondOnSExplicit})--(\ref{eq:HypVar2}) yield (\ref{eq:N=1eqsAppendix1}), (\ref{eq:N=1eqsAppendix2}), while (\ref{eq:GauginoEqOriginal}) is (\ref{eq:N=1eqsAppendix3}). Hence we just need to prove that (\ref{eq:HypVar3}), (\ref{eq:HypVar4}) can be rewritten as (\ref{eq:N=1eqsAppendix4}). As a first thing, we use in turn the following identities of special K\"ahler geometry
\be\nnb
(\im \mathcal G)^{-1\,IJ}  =  -(\im \cl N_1)^{-1\,IJ} -2e^{K_1}(Z^I\bar Z^J + \bar Z^IZ^J)\,,
\ee
\be\nnb
-(\im\cl N_1)^{-1\:IJ}=  2e^{K_1}\big(D_kZ^I g^{k\bar{l}}  D_{\bar{l}}\bar{Z}^J + \bar{Z}^I Z^J\big)
\ee
to rewrite
\be\nnb
P_I(\im \mathcal G)^{-1\,IJ}\; = \; 2e^{K_1} P_I\big(D_kZ^I g^{k\bar{l}}  D_{\bar{l}}\bar{Z}^J - Z^I\bar Z^J \big).
\ee
Next, recalling the definition of $P_I^{\;\ul j}$ given below (\ref{eq:HypVar4}), we observe that $P_I Z^I = 0$ and that
$P_I D_k Z^I = P_I \delta^I_k = e_k^{\ul j}\,.$
Hence (\ref{eq:HypVar3}), (\ref{eq:HypVar4}) are equivalent to (provided $a$ and $b$ do not vanish, which is guaranteed by (\ref{eq:N=1eqsAppendix1}) once one fixes $\mu\neq 0$)
\be\nnb
D_{\bar \imath}\bar Z^J \big(Q_{J\bb A} - \mathcal N_{1\,JK} Q^K_{\;\;\,\bb A}  \big) \Pi_2^{\bb A} \;\;=\;\; 0 \;\;=\;\; D_{i}Z^J \big(Q_{J\bb A} - \overline{\mathcal N}_{1\,JK} Q^K_{\;\;\,\bb A}  \big) \Pi_2^{\bb A}\,.
\ee
Recalling that in special K\"ahler geometry $ D_{i} Z^J \overline{\cl N}_{1\,JK} = D_i \mathcal G_K\,,$
we arrive at
\be\nnb
\Pi_2^T Q \,\C_1 D_i \Pi_1 \;=\; 0\; =\; \Pi_2^T Q\, \C_1D_{\bar \imath} \overline\Pi_1\,,
\ee
which is precisely the content of (\ref{eq:N=1eqsAppendix4}).

\subsection{In a symplectically covariant algebraic form}\label{SymplCovForm}

Continuing to revisit the analysis of \cite[sect.\:4]{Cassani:2007pq}, in the following we show that the $\cl N=1$ supersymmetry conditions (\ref{eq:N=1eqsAppendix1})--(\ref{eq:N=1eqsAppendix4}) can be reformulated in a symplectically covariant algebraic form as
\bea
\label{eq:4.28CasBil} \Pi_2^T Q  - 2 \hat{\bar\mu}e^{- \varphi} \re\big(e^{i\gamma}\Pi_1^T\big) &=& 0\\ [2mm]
\label{eq:4.35CasBil} Q\bb C_1\re\big( e^{i\gamma} \Pi_1 \big) &=&0\\ [2mm]
\label{eq:4.36CasBil} 2Q\bb C_1 \im \big( e^{i\gamma} \Pi_1 \big) - 6 \im\big( \hat\mu e^{-\varphi} \bb C_2\Pi_2 \big) - e^{\varphi}\bb M_2 (c+ \widetilde Q \xi) & = & 0\,.
\eea
Notice that, provided $\hat \mu\neq 0$, eq.\ (\ref{eq:4.35CasBil}) is actually implied by (\ref{eq:4.28CasBil}), upon multiplication of the latter by $Q\C_1$ and use of constraint (\ref{const}). In the main text, we employ a generalization of (\ref{eq:4.28CasBil})--(\ref{eq:4.36CasBil}) to study the extremization of the scalar potential $V$.

In order to derive (\ref{eq:4.28CasBil}), we multiply the two equations in (\ref{eq:N=1eqsAppendix4}) respectively  by  $g^{i\bar \jmath} D_{\bar\jmath} \overline\Pi{}_1^{\,T}$ and $g^{\bar\imath j} D_j \Pi_1^T$. Recalling (\ref{eq:KillingPrep}) and the special K\"ahler identity (\ref{skind1}), we get
\bea
\nnb \Pi_2^T Q \,\C_1\,  \big(\ft12 \C_1^T \bb M_1\C_1 +\ft i2 \bb C_1 + \overline\Pi_1\Pi_1^T \big) &=& 0\\ [2mm]
\label{eq:VarHyperIntermediate}\Pi_2^T Q \,\C_1\,  \big(\ft12 \C_1^T \bb M_1\C_1 -\ft i2 \bb C_1 + \Pi_1\overline \Pi_1^T \big) &=& 0\,.
\eea
Subtracting the second from the first, we have
\be
\nnb 2\Pi_2^T Q  +i e^{-\varphi}\big(\cl P_-\Pi_1^T - \cl P_+ \overline\Pi{}_1^T \big) \; = \; 0\,,
\ee
which yields (\ref{eq:4.28CasBil}) upon use of (\ref{eq:N=1eqsAppendix2}). One can also see that summing up the two conditions (\ref{eq:VarHyperIntermediate}) the same equation is retrieved (the identity $\bb M_1 \Pi_1=-i \C_1\Pi_1 $ is required in the computation). We also checked that in the case of the universal hypermultiplet, where (\ref{eq:N=1eqsAppendix4}) does not hold, eq.\ (\ref{eq:4.28CasBil}) follows from (\ref{eq:N=1eqsAppendix2}) alone.

To derive (\ref{eq:4.35CasBil}), (\ref{eq:4.36CasBil}) we start from the susy condition $D_{a} ( \cl P_3 \mp e^{\pm i\gamma} \cl P_\pm)=0$, and multiply it by  $g^{a \bar b} D_{\bar b} \overline \Pi_{2}$. We obtain
 \bea
 0 &=& g^{a \bar b}D_{\bar b} \overline \Pi_{2} D_{a} \Pi{}_2^T   \big[ e^{\varphi} \C_2(c+\widetilde Q \xi)\,\mp\, 2e^{\pm i\gamma} Q \C_1 (\re\Pi_1 \pm  i\im\Pi_1 ) \big]  \nn\\ [2mm]
\nnb &=& \left(-\ft12  \C_2^T \bb M_2 \C_2+ \ft i2 \C_2 - \Pi_2 \overline\Pi{}_2^{\,T}\right) \big[ e^{\varphi} \C_2(c+\widetilde Q \xi)\,\mp\,2 e^{\pm i\gamma } Q \C_1 (\re\Pi_1 \pm  i\im\Pi_1 )\big] \\ [2mm]
\nnb &=& \ft12\left(\C_2^T \bb M_2  - i\one \right)\big[ e^{\varphi} (c+\widetilde Q \xi)\,\pm\,  2e^{\pm i\gamma}  \C_2Q \C_1 (\re\Pi_1 \pm  i\im\Pi_1 ) \big] -  3i\hat \mu e^{- \varphi}  \Pi_2\,,
\eea
where to get the second line we use identity (\ref{skind2}), while the third line is obtained recalling (\ref{eq:N=1eqsAppendix2}). Adding up these two equations and taking the real part we arrive to (\ref{eq:4.36CasBil}); no further informations is contained in the imaginary part, since one can check that it is just (\ref{eq:4.36CasBil}) multiplied by $\bb M_2$. Analogously, taking either the imaginary or the real part of the difference of the two equations above we get (\ref{eq:4.35CasBil}).

\vskip 3mm

Finally, we remark that in turn (\ref{eq:4.28CasBil})--(\ref{eq:4.36CasBil}) imply (\ref{eq:N=1eqsAppendix2})--(\ref{eq:N=1eqsAppendix4}), and are therefore {\it equivalent} to them. Indeed, eq.\ (\ref{eq:N=1eqsAppendix2}) is obtained by contracting eq.\ (\ref{eq:4.28CasBil}) with $\C_1\Pi_1$ as well as with $\C_1\overline\Pi{}_1$, and eq.\ (\ref{eq:4.36CasBil}) with $\Pi_2^T$. Contraction of (\ref{eq:4.35CasBil}), (\ref{eq:4.36CasBil}) with $D_a\Pi_2^T$ yields (\ref{eq:N=1eqsAppendix3}), while multiplication of (\ref{eq:4.28CasBil}) by $\C_1 D_i\Pi_1$ or by $\C_1 D_{\bar\imath}\overline\Pi_1$ provides (\ref{eq:N=1eqsAppendix4}). The following special K\"ahler geometry relations are needed in the proof:
\bea
\nnb &&\Pi^T_2 \C_2D_a \Pi_2 \;=\; 0\;=\; \overline\Pi{}^{\,T}_2 \C_2D_a \Pi_2\qquad\quad (\textrm{same with }a\to i\;,\;2\to 1)\\ [2mm]
\nnb &&\bb M_2 \Pi_2 = -i\,\bb \C_2\Pi_2\qquad ,\qquad \bb M_2 D_a\Pi_2 = +i\,\bb \C_2D_a\Pi_2\,.
\eea

\section{\!Flux/gauging dictionary for IIA \hbox{on SU(3)$\:$structure}}\label{ScalPotIIA}

Gauged $\cl N=2$ supergravities with a scalar potential of the form
studied in this paper can be derived by flux compactifications of
type II theories on $SU(3)$ and $SU(3)\times SU(3)$ structure
manifolds. While we refer to the literature (see e.g.
\cite{Gurrieri:2002wz, Gurrieri:2002iw, Tomasiello:2005bp, GLW1, KashaniPoor:2006si, GLW2, Cassani:2007pq, House:2005yc, KashaniPoorNearlyKahler,
Cassani:2008rb, Cassani:2009ck, Grana:2009im}) for a detailed study of such general $\cl
N=2$ dimensional reductions and the related issues,\footnote{In this
context, see also \cite{Benmachiche:2006df, Koerber:2007xk,
Martucci:2009sf, Derendinger:2004jn, Villadoro:2005cu, Micu:2007rd} for studies
of compactifications preserving $\cl N=1$.} in this appendix we
provide a practical dictionary between the 10d and the 4d
quantities, with a focus on the scalar potential derived from $SU(3)$
structure  compactifications of type IIA. In particular, we
illustrate how the expressions one derives for $V_{\rm NS}$ and
$V_{\rm R}$ are consistent with the scalar potential (\ref{pot0})
studied in the main text.

\subsection{SU(3) structures and their curvature}\label{sfluxsu3}

An $SU(3)$ structure on a 6d manifold $M_6$ is defined by a real
2--form $J$ and a complex, decomposable\footnote{A $p$--form is
decomposable if locally it can be written as the wedging of $p$
complex 1--forms.} 3--form $\Omega$, satisfying the compatibility
relation $J\wedge\Om = 0$ as well as the non-degeneracy (and
normalization) condition \be\label{eq:NormalizationJOm} \ft
i8\Om\wedge \bar\Om \;=\; \ft 16 J\wedge J\wedge J \;=\; vol_6
\;\neq\; 0 \quad{\rm everywhere}\,. \ee $\Om$ defines an almost
complex structure $I$, with respect to which is of type $(3,0)$. In
turn, $I$ and $J$ define a metric on $M_6$ via $g= JI$. The latter
is required to be positive-definite, and $vol_6$ above denotes the
associated volume form.

$SU(3)$ structures are classified by their torsion classes
$W_i\,,\,i=1,\ldots 5$, defined via \cite{Chiossi:2002tw}: \bea \nnb
dJ &=& \ft 32\im(\ol W_1 \Om) +  W_4\wedge J + W_3\\ [2mm]
\label{eq:SU3torsion} d\Om &=& W_1\wedge J\wedge J + W_2\wedge J
+\ol W_5\wedge \Om\,, \eea where $W_1$ is a complex scalar, $W_2$ is
a complex primitive (1,1)--form (primitive means $W_2\wedge J\wedge
J=0$), $W_3$ is a real primitive (1,2) + (2,1)--form (primitive
$\Leftrightarrow W_3\wedge J=0$), $W_4$ is a real 1--form, and $W_5$
is a complex (1,0)--form.

Ref. \cite{BedulliVezzoni} provides a formula for the Ricci scalar
$R_6$ in terms of the torsion classes. We will restrict to
$W_4=W_5=0$, in which case the formula is \be R_6\;=\;\ft 12\big(
15|W_1|^2 -W_2\lrcorner\overline W_2 - W_3\lrcorner W_3 \big)\,. \ee
This can equivalently be expressed as \be\label{eq:RInTermsOfJandOm}
R_6 \,vol_6 = -\ft 12\big[\, dJ\wedge *dJ + d\Om \wedge *d\bar\Om -
(dJ \wedge \Om)\wedge *(dJ\wedge \bar \Om) \,\big]\,, \ee as it can
be seen recalling (\ref{eq:SU3torsion}) and computing \bea \nnb
d\Om\wedge *d\bar\Om &=& 12|W_1|^2 vol_6 - J\wedge W_2 \wedge
\overline W_2 \;=\; \big(12|W_1|^2 + W_2\lrcorner\overline
W_2\big)vol_6  \\ [1mm] \nnb dJ\wedge *dJ &=& \big(9 |W_1|^2 +
W_3\lrcorner W_3\big)vol_6 \\ [1mm] \!\!\!\!\!\!(dJ\wedge \Om
)\wedge *(dJ \wedge \bar \Om)&=& 36|W_1|^2 vol_6\,. \eea

\subsection{The scalar potential from dimensional reduction}

The 4d scalar potential receives contributions from both the NSNS
and the RR sectors of type IIA supergravity. These are respectively
given by \bea \nnb \!\!\!\!\!\! V_{\mathrm{\rm NS}}  &=&
\frac{e^{2\varphi}}{2\mathscr V}\int_{M_6} \big(\,\ft 12 H\wedge * H \,
-R_6 *1\,  \big)\\ [2mm]
\label{eq:DefV_NS} &=&\frac{e^{2\varphi}}{4\mathscr V}\int_{M_6} \Big[\,H\wedge *H + dJ\wedge *dJ +   d\Om \wedge *d\bar\Om - (dJ \wedge \Om)\wedge *(dJ\wedge \bar \Om)\, \Big],\\[2mm]
\label{eq:Vrr} \!\!\!V_{\rm R}\; &=& \;
\frac{e^{4\varphi}}{4}\int_{M_6}\big(\,F_0^2 *1 \,+\,  F_2\wedge
*F_2 \,+\, F_4\wedge *F_4 \,+\, F_6\wedge *F_6\,\big), \eea and the
total potential reads $V=V_{\rm NS} + V_{\rm R}$. In
(\ref{eq:DefV_NS}), $H$ is the internal NSNS field-strength, $\mathscr V
= \int_{M_6}vol_6$, and $\varphi$ is the 4d dilaton $e^{-2\varphi}=
e^{-2\phi}\mathscr V$, where we are assuming that the 10d dilaton $\phi$
is constant along $M_6$. The $k$--forms $F_k$ appearing in
expression (\ref{eq:Vrr}) are the internal RR field strengths,
satisfying the Bianchi identity $dF_k-H\wedge F_{k-2}=0$. The $F_6$
form can be seen as the Hodge-dual of the $F_4$ extending along
spacetime, and the term $F_6\wedge *F_6$ arises in a natural way if
one considers type IIA supergravity in its democratic formulation
\cite{Bergshoeff:2001pv}.

\subsubsection*{Expansion forms}

In order to define the mode truncation, we postulate the existence
of a basis of differential forms on the compact manifold in which to
expand the higher dimensional fields. For a detailed analysis of the
relations that these forms need to satisfy in order that the
dimensional reduction go through, see in particular
\cite{KashaniPoor:2006si}.

We take $\om_0=1$ and $\tilde\om^0= \frac{vol_6}{\mathscr V}$, and we
assume there exist a set of 2--forms $\om_a$ satisfying
\be\label{gabFromom} \,\omega_a \wedge * \omega_b\,=\, 4\,
g_{ab}\,vol_6\qquad,\qquad\om_a\wedge \om_b = -d_{abc} \tilde\om^c,
\ee\vskip -1mm\noindent where $g_{ab}$ should be independent of the
internal coordinates, $d_{abc}$ should be a constant tensor, and the
dual 4--forms $\tilde \om^a$ are defined as \be \tilde \om^a = -\ft
1{4\mathscr V}g^{ab}*\om_b\,. \ee \vskip -1mm\noindent From the above
relations, we see that \be \om_a\wedge \tilde \om^b \,=\,
-\delta_a^b\, \tilde\om^0\qquad ,\qquad \om_a\wedge\om_b\wedge\om_c
= d_{abc}\tilde\om^0\,. \ee\vskip -1mm\noindent We also assume the
existence of a set of 3--forms $\alpha_I,\beta^I$, satisfying \be
\alpha_I\wedge \beta^J \,=\, \delta_I^J\, \tilde\om^0. \ee\vskip
-1mm Adopting the notation $\omega^{\bb A}=(\tilde
\omega^A,\omega_A)^T =(\tilde \omega^0,\tilde \omega^a,
\omega_0,\omega_a)^T\;$ and $\; \alpha^{\bb I}=(\beta^I,
\alpha_I)^T$, we see that the symplectic metrics $\bb C$ appearing
in the main text are here given by \be \C_1^{\bb I \bb J}\,=\, -\int
\alpha^{\bb I} \wedge  \alpha^{\bb J} \qquad,\qquad \C_2^{\bb A \bb
B} \,=\,-\int\langle  \omega^{\bb A}, \omega^{\bb B}\rangle\,,
\ee\vskip -1mm\noindent where the antisymmetric pairing
$\langle\,,\,\rangle$ is defined on even forms $\rho,\sigma$ as
$\langle \rho,\sigma\rangle\, =\, [\lambda(\rho)\wedge
\sigma]_{\mathrm{6}}$, with $\,\lambda(\rho_{k}) =
(-)^{\frac{k}{2}}\rho_{k}\,$, $\,k$ being the degree of $\rho$, and
$[\;\;]_6$ selecting the piece of degree 6.

The basis forms are used to expand $\Om$ as \be\Omega \,=\, Z^I
\alpha_{ I} -{\cal G}_I \beta^I \,=\, e^{-\frac{K_1}{2}}\Pi_1^{\bb
I} \,  \alpha_{\bb I}\,, \ee and $J$ together with the internal NS
2--form $B$ as: \be J\,=\, v^a \om_a\;\;,\;\; B\, =\,
b^a\om_a\quad\; \Rightarrow \;\quad e^{-B-iJ}\,=\, X^A\om_A - \cl
F_A \tilde\om^A \,=\, e^{-\frac{K_2}{2}}\Pi_2^{\bb A}\om_{\bb A}\,,
\ee where in the last equalities we define $\alpha_{\bb I} = \bb
C_{\bb I\bb J}\alpha^{\bb J} = (\alpha_I, -\beta^I)^T$ and $\om_{\bb
A}= \bb C_{\bb A\bb B} \om^{\bb B}= (\om_A,-\tilde\om^A)^T$, and we
adopt the symplectic notation defined in (\ref{eq:SymplSections}).
Here, $(Z^I, \cl G_I)$ and $(X^A,\cl F_A)$  represent the
holomorphic sections on the moduli spaces of $\Om$ and $B+iJ$
expanded as above, which (under some conditions
\cite{GLW1,KashaniPoor:2006si,GLW2}) indeed exhibit a special
K\"ahler structure, and correspond respectively to the manifolds
$\mathscr M_1$ and $\mathscr M_2$ of the main text. Notice that here
$X^A \equiv (X^0,X^a) \equiv (1,x^a) = (1,-b^a-iv^a) $, while $\cl
F_A = \pd{\cl F}{X^A}$, where the cubic holomorphic function $\cl F
= \frac{1}{6}d_{abc}\frac{X^aX^bX^c}{X^0}$
is identified with the prepotential on $\mathscr M_2$. The K\"ahler
potentials on $\mathscr M_1$ and $\mathscr M_2$ are recovered from
$K_1 =-\log i\int\Om\wedge\bar \Om\,$ and $\,K_2 = -\log \ft 43 \int
J\wedge J \wedge J$, the latter yielding the metric $g_{ab}$
appearing in (\ref{gabFromom}). Notice that
(\ref{eq:NormalizationJOm}) implies $e^{-K_1} = e^{-K_2} = 8\mathscr V$.

The matrices $\bb M$ defined in (\ref{eq:DefbbM}) are given by
\be\label{eq:bbMfromForms} \bb M_{1,\bb I \bb J}\,=\, -\int
\alpha_{\bb I} \wedge  *\alpha_{\bb J} \qquad,\qquad \bb M_{2,\bb A
\bb B}\,=\,-\sum_k\int (e^B\omega_{\bb A})_k\wedge  *(e^B\omega_{\bb
B})_k\,, \ee and from the second relation one finds that the period
matrix $\cl N_2$ on $\mathscr M_2$ reads
\be\label{eq:ReNImN} \!\!\!\!\!\!\!\!\!\re\cl N_{AB} =
-\left(\begin{array}{cccc} \frac{1}{3} d_{abc}b^ab^bb^c &
\frac{1}{2}d_{abc}b^bb^c \\ [2mm] \frac{1}{2}d_{abc}b^bb^c &
d_{abc} b^c
\end{array} \right)\;,\qquad
\im\cl{N}_{AB} = -4\mathscr V \left( \begin{array}{cc} \ft 14
+g_{ab}b^ab^b & g_{ab} b^b\\ [2mm] g_{ab} b^b &  g_{ab}
\end{array} \right),
\ee which is in agreement with the expression derived from $\cl F$
via the standard formula \cite{Craps:1997gp} \be \cl {\cal
N}_{AB}\;=\; \overline{\cl F}_{AB} + 2i {\textstyle{\frac{\im \cl F_{AD}X^D\im
\cl F_{BE}X^E}{X^C\im \cl F_{CE}X^E} }}\;,\qquad\quad {\textstyle{\cl
F_{AB}\equiv\frac{\partial^2\cl F}{\partial X^A\partial X^B} }}\,. \ee
Finally, we also require the following differential conditions on
the basis forms: \be\label{eq:DiffCondSU3} d\om_a \,=\, e_a{}^{\bb
I}\alpha_{\bb I}\qquad,\qquad d\alpha^{\bb I} \,=\, e_a{}^{\bb I}
\tilde\om^a\qquad,\qquad d\tilde\om^a\,=\, 0\,, \ee where the
$e_a{}^{\bb I}= (e_a{}^I,e_{aI})$ are real constants, usually called
`geometric fluxes'. Defining the total internal NS 3--form as $H=
H^{\rm fl} +dB$, and  expanding its flux part as \be H^{\rm fl}
\,=\, -e_0{}^I\alpha_I + e_{0I}\beta^I \,\equiv\, -e_0{}^{\bb
I}\alpha_{\bb I}\,, \ee with constant $e_0{}^{\bb I}$, we can define
$e_{A}{}^{\bb I} = (e_0{}^{\bb I},e_a{}^{\bb I})^T$, and thus fill
in half of the charge matrix $Q$ introduced in (\ref{DefCharges}):
\be\label{eq:HalfCharges} Q_{\bb A}{}^{\bb I} =
\left(\begin{array}{c}  e_{A}{}^{\bb I}\\ 0\end{array}\right). \ee
As first noticed in \cite{GLW2}, more general matrices, involving the
$m_A{}^{\bb I}$ charges as well, can be obtained by considering
non-geometric fluxes, or $SU(3) \times SU(3)$ structure
compactifications. The nilpotency condition $d^2=0$ applied to
(\ref{eq:DiffCondSU3}), together with the Bianchi identity $dH=0$,
translates into the constraint \be e_A{}^{\bb I}e_{B\bb I}
\,=\,0\qquad\qquad\qquad \textrm{with }\;\; e_{A \bb
I}=\mathbb{C}_{\bb I \bb J} e_{A}{}^{\bb J}, \ee which, taking into
account (\ref{eq:HalfCharges}), is consistent with (\ref{const}).

In the following, by using the above  relations we recast in turn
expressions (\ref{eq:DefV_NS}), (\ref{eq:Vrr}) for $V_{\rm NS}$ and
$V_{\rm R}$ in terms of 4d degrees of freedom, and show their
consistency with (\ref{pot0}).

\subsubsection*{Derivation of $V_{\rm NS}$}

Recalling the expansions of $J$, $H$ and $\Om$ defined above, using
the assumed properties of the basis forms, and adopting the notation
introduced in (\ref{eq:SymplSections}), one finds \bea \nnb \int
dJ\wedge *dJ  =- v^av^b\,  e_a{}^{\bb I}\, \mathbb{M}_{1,\bb I \bb
J} e_{b}{}^{\bb J} \quad &,& \quad \int H\wedge *H = -b^Ab^B \,
e_A{}^{\bb I} \, \mathbb{M}_{1,\bb I \bb J} e_{B}{}^{\bb J}\,, \\
[2mm] \nnb \int d\Om\wedge *d\bar\Om  =
{\textstyle{\frac{e^{-K_1}}{4{\mathscr V}}}} \Pi_1^{\bb I} e_{a\bb I}
g^{ab}  e_{b\bb J}\ol \Pi{}_1^{\bb J} \quad &,& \quad\int(dJ \wedge
\Om)\wedge *(dJ\wedge \bar \Om) = {\textstyle{\frac{e^{-K_1}}{\mathscr
V}}} \Pi_1^{\bb I} e_{a\bb I} v^a v^b e_{b\bb J}\ol \Pi{}_1^{\bb J},
\eea where we define $b^A = (-1,b^a)$. Plugging this into
(\ref{eq:DefV_NS}), we get the NSNS contribution to $V$, expressed
in a 4d language: \be\label{eq:VNSfinal} V_{\mathrm{\rm NS}} =
-\frac{e^{2\varphi}}{4\mathscr V} \Big[ X^{A} e_A{}^{\bb I}\,
\mathbb{M}_{1,\bb I \bb J} e_{B}{}^{\bb J} \, \ol X{}^{B} -
 {\textstyle{\frac{e^{-K_1}}{4\mathscr V} }} \Pi_1^{\bb I} e_{a\bb I} (g^{ab} - 4v^av^b) e_{b\bb J}\ol \Pi{}_1^{\mathbb J}  \Big]\,.
\ee Recalling (\ref{eq:ReNImN}), noticing that $\ft 1{4\mathscr V}
(g^{ab}-4v^av^b) = -(\im \cl N_2)^{-1\,ab}-4e^{K_2}(X^a\ol X{}^b +
\ol X{}^a X^b)$, and recalling that $e^{-K_1} = e^{-K_2} = 8\mathscr V$,
we conclude that (\ref{eq:VNSfinal}) is consistent with
(\ref{pot0}).

\subsubsection*{Derivation of $V_{\rm R}$}

We consider the internal field-strength $G = G_0+G_2+G_4+G_6$, defined as  $F_k =
\sqrt 2\big(e^{B}G\big)_k$. The $G_k$ satisfy the Bianchi identity $dG_k -
H^{\rm fl}\wedge G_{k-2} = 0$. We define the expansion
\be \nnb
G_0 \,=\, p^0\qquad\; ,\;\qquad G_2 \,=\, p^a\om_a\qquad \;,\;\qquad
A_3 \,=\, \xi^{\bb I} \alpha_{\bb I} \ee \be\nnb \!\! G_4 \,=\,
G_4^{\rm fl} + dA_3 \,=\, -(q_{a} - e_{a\bb I}\xi^{\bb I} )
\tilde\om^a\qquad,\qquad G_6\,=\, G_{6}^{\rm fl} - H^{\rm fl} \wedge
A_3 \,=\,    -(q_{0} - e_{0\bb I}\,\xi^{\bb I})  \tilde\om^0, \ee
where $p^0,p^a,q_0,q_a$ are constant, while the $\xi^{\bb I}$ are 4d scalars. The
Bianchi identities then amount just to the following constraint
among the charges \be\label{eq:ConstrFluxesSU3} p^A e_A{}^{ \bb I}
\, =\, 0\,, \ee which, recalling (\ref{eq:HalfCharges}), gives the
last equality in (\ref{const}). Then the integral in (\ref{eq:Vrr})
reads \be \ft 12\sum_k \int F_k \wedge * F_k \,=\, \sum_k \int (e^BG)_k
\wedge * (e^BG)_k\,=\, (c + \widetilde Q \xi)^T\bb M_2 (c +
\widetilde Q \xi)\,, \ee \vskip -3mm\noindent where for the second
equality we use (\ref{eq:bbMfromForms}), and here $(c + \widetilde Q
\xi)^{\bb A} \, =\, (p^A, \, q_{A} - e_{A\bb I}\xi^{\bb I} )^T$. The
expression for $V_{\rm R}$ we obtain is therefore consistent with
(\ref{pot0}).


\section{Details on U-invariance\label{U-Invariance}}
\label{sAd}


The explicit expression of the quartic $G$-invariant associated to a $d$-special K\"{a}hler space~$G/H$ is \cite{QuarticInv}
\begin{eqnarray}
&& \mathcal{I}_{4}\left( c^{4}\right)  =    d_{\mathbb{A}_{1}\mathbb{A}_{2}%
\mathbb{A}_{3}\mathbb{A}_{4}}c^{\mathbb{A}_{1}}c^{\mathbb{A}_{2}}c^{\mathbb{A%
}_{3}}c^{\mathbb{A}_{4}}, \label{expli} \nn\\
&&= 6d_{~~00}^{00}q_{0}^{2}\left( p^{0}\right)
^{2} +4d_{0}^{~abc}p^{0}q_{a}q_{b}q_{c} +6d_{~~cd}^{ab}q_{a}q_{b}p^{c}p^{d} 
+24d_{~0~b}^{0~a}q_{0}p^{0}q_{a}p^{b} +4d_{~abc}^{0}q_{0}p^{a}p^{b}p^{c}\nn\\ 
&&= -(p^0 q_0+p^a q_a)^2 +\ft{2}{3}d_{abc}\,q_{0}p^{a}p^{b}p^{c}- 
\ft{2}{3}d^{abc}\,p^{0} q_{a}q_{b}q_{c} +d_{abc}d^{aef}p^{b}p^{c}q_{e}q_{f} 
\,, \nn \end{eqnarray}
with $d_{\mathbb{A}_{1}\mathbb{A}_{2}\mathbb{A}_{3}\mathbb{A}_{4}}=d_{(\mathbb{A}%
_{1}\mathbb{A}_{2}\mathbb{A}_{3}\mathbb{A}_{4})}$
throughout. 
Thus, the characterization of $\mathcal{I}_{4}\left( c^{4}\right) $ as an $%
Sp\left( 2h_{2}+2,\mathbb{R}\right) $-scalar entails that the unique
independent non-vanishing components of the corresponding $d_{\mathbb{A}_{1}\mathbb{A}_{2}%
\mathbb{A}_{3}\mathbb{A}_{4}}$ read as follows:
\begin{eqnarray}
d_{~~00}^{00} \!&=&\!-\frac{1}{6}\;,\qquad\qquad
d_{0}^{~abc} \;\equiv \;-\frac{1}{6}d^{abc}\,,\qquad\qquad
d_{~~cd}^{ab} \;\equiv \;\frac{1}{6}\left( d_{ecd}d^{eab}-\delta
_{(c}^{a}\delta _{d)}^{b}\right),\nn \\ [2mm]
d_{~0~b}^{0~a}\!&=&\!-\frac{1}{12}\delta _{b}^{a}\;,\qquad\qquad
d_{~abc}^{0} \;\equiv \;\frac{1}{6}d_{abc}\,,\label{uinv-5}
\end{eqnarray}
where $d_{abc}=d_{\left( abc\right) }$ and $d^{abc}=d^{\left( abc\right) }$
are the covariant and contravariant $d$-tensor defining the $d$-special K%
\"{a}hler geometry of vector multiplets' scalar manifold
$G/H$. Notice that, whereas $d_{abc}$ is always
scalar-independent, $d^{abc}$ is
generally scalar-dependent. Nevertheless, (\textit{at least}) in \textit{%
symmetric} $d$-special K\"{a}hler geometries $d^{abc}$ is
scalar-independent, and thus so is $\mathcal{I}_{4}\left(
c^{4}\right) $.
The completely symmetric tensor $d_{\mathbb{A}_{1}\mathbb{A}_{2}\mathbb{A}%
_{3}\mathbb{A}_{4}}$ whose unique independent non-vanishing
components are given by (\ref{uinv-5}) is the
unique \textit{invariant} rank-$4$ tensor of the repr.
$\mathbf{R}_{G}$ of $G$. An identical argument
can be used for the unique invariant rank-$4$ tensor $d_{\mathbb{I}%
_{1}\mathbb{I}_{2}\mathbb{I}_{3}\mathbb{I}_{4}}$ of the repr. $\mathbf{R}_{%
\mathscr{G}}$ of $\mathscr{G}$, defined in terms of the tensors
$d_{ijk}$
and $d^{ijk}$ determining the $d$-special K\"{a}hler geometry of $
\mathscr{G}/\mathscr{H}$, whose the hypermultiplets' scalar manifold ${%
\mathscr M}_{Q}$ is the $c$-map.

We now detail computations of various quantities, generally
covariant with respect to $\mathbf{R}_{G}\times
\mathbf{R}_{\mathscr{G}}$, useful in the
treatment given in subsection \ref{U-Invariant-Lambda}. Firstly, by using the relation (\ref{idsym}) (holding \textit{at
least} in \textit{homogeneous symmetric} $d$-special K\"{a}hler
geometries), the constraint (\ref{13}) implies
$\mathcal{I}_{4}\left( c^{4}\right) $ to read  
\be
\mathcal{I}_{4}\left( c^{4}\right)  \;=\;-\left( p^{0}\right) ^{2}\mathbf{q}%
_{0}^{2}\quad \qquad\qquad\qquad\textrm{with}\qquad \mathbf{q}_{0} \equiv q_{0}+\frac{1}{6}\frac{d_{abc}p^{a}p^{b}p^{c}}{%
\left( p^{0}\right) ^{2}}.
\ee
  Now let us consider the  invariant ${\cal I}_{16}(c^4 Q^{12})$. 
  \begin{eqnarray}
\mathcal{I}_{16}\left( c^{4}Q^{12}\right) &\equiv &d_{\mathbb{I}_{1}\mathbb{I%
}_{2}\mathbb{I}_{3}\mathbb{I}_{4}}d_{\mathbb{I}_{5}\mathbb{I}_{6}\mathbb{I}%
_{7}\mathbb{I}_{8}}d_{\mathbb{I}_{9}\mathbb{I}_{10}\mathbb{I}_{11}\mathbb{I}%
_{12}}d_{\mathbb{A}_{1}\mathbb{B}_{1}\mathbb{B}_{2}\mathbb{B}_{3}}d_{\mathbb{%
A}_{2}\mathbb{B}_{5}\mathbb{B}_{6}\mathbb{B}_{7}}d_{\mathbb{A}_{3}\mathbb{B}%
_{9}\mathbb{B}_{10}\mathbb{B}_{11}}d_{\mathbb{A}_{4}\mathbb{B}_{4}\mathbb{B}%
_{8}\mathbb{B}_{12}}  \notag \\ [1mm]
&&c^{\mathbb{A}_{1}}c^{\mathbb{A}_{2}}c^{\mathbb{A}_{3}}c^{\mathbb{A}_{4}}Q^{%
\mathbb{B}_{1}\mathbb{I}_{1}}\ldots 
Q^{\mathbb{B}_{12}\mathbb{I}_{12}} .
\label{I_16-our-case-def2}
\end{eqnarray}
We set all components of $Q^{\mathbb{AI}}$, $c^{\bb A}$  to zero 
except for 
\be
  Q_{a}^{~i} = -e_{a}{}^i \qquad ,\qquad  Q_{a0}=-e_{a} \qquad,  \qquad c^0=p^0 \qquad ,\qquad c_0={\bf q_0} \,.
\ee
With this choice the only contributions come from
   the components $d^0{}_{ijk}=-\ft16 \, d_{ijk}$ of 
$d_{\bb I_1 .. \bb I_4}$ and $d_0{}^{b_1 b_2 b_3}$ of $d_{\bb A\, \bb B_1 \bb B_2 \bb B_3}$ .  
 More precisely, one can write  
 \be
\!\!\!\!\! \mathcal{I}_{16}\left( c^{4}Q^{12}\right) ={1\over 6^4} (p^0)^4 \,d_{\mathbb{I}_{1}\mathbb{I%
}_{2}\mathbb{I}_{3}\mathbb{I}_{4}}d_{\mathbb{I}_{5}\mathbb{I}_{6}\mathbb{I}%
_{7}\mathbb{I}_{8}}d_{\mathbb{I}_{9}\mathbb{I}_{10}\mathbb{I}_{11}\mathbb{I}%
_{12}}d^{ b_1 b_2 b_3  } d^{ b_5 b_6 b_7  }d^{ b_9 b_{10} b_{11}  }d^{ b_4 b_8 b_{12}  }
 Q_{b_1}{}^{ \mathbb{I}_{1}}\ldots 
Q_{b_{12}}{}^{\mathbb{I}_{12}}\,.\nn\\
\ee
 There are four different type of  contributions  depending on how indices are contracted.
 They are given by  ${1\over 6^7} \,(p^0)^4\, \beta^3\, \times $
 \bea
\!\!\!\!\!\!\!&& D^3 d^{b_4 b_8 b_{12}} e_{b_4}  e_{b_8} e_{b_{12}} \,=\, 6  \, D^3  R \label{4terms} \\ [1mm]
\!\!\!\!\!\!\!&& 9\, D^2 \,d_{b_2  b_3 b_4}  \, d^{b_1 b_2 b_3}\, d^{ b_4 b_8 b_{12}  }  \,e_{b_1}  e_{b_8} e_{b_{12}}  \,=\,18   \, D^2 \, (h+3) R  \nn\\ [1mm]
\!\!\!\!\!\!\!&& 27 \,D\, d_{b_2  b_3 b_4} \, d_{b_6  b_7 b_8} \, d^{b_1 b_2 b_3}\, d^{ b_5 b_6 b_7  }\, d^{ b_4 b_8 b_{12}  }  \,e_{b_1}  e_{b_5} e_{b_{12}}   
\,=\, 18 \, D \, (h+3)^2 R  \nn\\ [1mm]
\!\!\!\!\!\!\!&& 27 \,D\, d_{b_2  b_3 b_4} \, d_{b_6  b_7 b_8} \,  d^{  b_{10} b_{11} b_{12}  }\, d^{b_1 b_2 b_3}\, d^{ b_5 b_6 b_7  }\, d^{ b_9 b_{10} b_{11}  }d\,d^{ b_4 b_8 b_{12}  }  \,e_{b_1}  e_{b_5} e_{b_{9}}   
\,=\,6\, (h+3)^3 R  \nn
 \eea
with $D=d^{abc} d_{abc}$, while $R$ was defined in (\ref{R-def}). In writing (\ref{4terms}) we use the symmetric $d$-special K\"ahler identity
 \begin{eqnarray}
 d_{abf}d^{a(bc}d^{de)f}=\frac{4}{3}\delta
_{b}^{(b}d^{cde)} =\ft13( h_{2}+3)d^{cde}.  \nn
\end{eqnarray}
Collecting all pieces together one finds
\be
{\cal I}_{16}(c^4 Q^{12})\,=\, {1\over 6^6} \,(p^0)^4 \beta^3 R\,(D+h_2+3 )^3.
\ee

\end{appendix}

\newpage
%
%

\providecommand{\href}[2]{#2}\begingroup\raggedright\endgroup

\end{document}